\documentclass[utf8]{FrontiersinHarvard}

\usepackage{url,hyperref,lineno,microtype,subcaption}
\usepackage[onehalfspacing]{setspace}
\usepackage{graphicx}
\usepackage{txfonts}
\usepackage{tabularx}
\usepackage{csvsimple}
\usepackage{booktabs}
\usepackage{longtable}
\usepackage{caption}
\usepackage{subcaption}
\usepackage{lscape}
\usepackage{appendix}
\usepackage{ulem} 
\usepackage{threeparttable}

\def\keyFont{\fontsize{8}{11}\helveticabold }
\def\firstAuthorLast{Sample {et~al.}} 
\def\Authors{Zhensen Fu\,$^{1,2}$, Zhaoxiang Qi\,$^{1,2,*}$, Shilong Liao\,$^{1,2,*}$, Xiyan Peng\,$^{1}$, Yong Yu\,$^{1,2}$, Qiqi Wu\,$^{1,2}$, Li Shao\,$^{3}$ and Youhua Xu\,$^{3}$}
\def\Address{$^{1}$Shanghai Astronomical Observatory, Chinese Academy of Sciences, 80 Nandan Road, Shanghai 200030, People’s Republic of China \\
$^{2}$University of Chinese Academy of Sciences, No.19(A) Yuquan Road, Shijingshan District, Beijing 100049, People’s Republic of China  \\
$^{3}$National Astronomical Observatory, Chinese Academy of Sciences, 20A Datun Road, Chaoyang District, Beijing 100012, People’s Republic of China}
\def\corrAuthor{Zhaoxiang Qi}

\def\corrEmail{zxqi@shao.ac.cn}

\begin{document}
\onecolumn
\firstpage{1}

\title[Running Title]{Simulation of CSST’s astrometric capability} 

\author[\firstAuthorLast ]{\Authors} 
\address{} 
\correspondance{} 

\extraAuth{Shilong Liao\\ shilongliao@shao.ac.cn}

\maketitle

\begin{abstract}
\section{}
The China Space Station Telescope (CSST) will enter a low Earth orbit around 2024 and operate for 10 years, with seven of those years devoted to surveying the area of the median-to-high Galactic latitude and median-to-high Ecliptic latitude of the sky. To maximize the scientific output of CSST, it is important to optimize the survey schedule. We aim to evaluate the astrometric capability of CSST for a given survey schedule and to provide independent suggestions for the optimization of the survey strategy. For this purpose, we first construct the astrometric model and then conduct simulated observations based on the given survey schedule. The astrometric solution is obtained by analyzing the simulated observation data. And then we evaluate the astrometric capability of CSST by analyzing the properties of the astrometric solution. We find that the accuracy of parallax and proper motion of CSST is better than 1 mas($\cdot$ yr$^{-1}$) for the sources of 18-22 mag in g band, and about 1$\sim$10 mas($\cdot$ yr$^{-1}$) for the sources of 22-26 mag in g band, respectively. The results from real survey could be worse since the assumptions are optimistic and simple. We find that optimizing the survey schedule can improve the astrometric accuracy of CSST. In the future, we will improve the astrometric capability of CSST by continuously iterating and optimizing the survey schedule. 
\tiny
 \keyFont{ \section{Keywords:} astrometry, data analysis, proper motion, parallax, simulations} 
\end{abstract}
\section{Introduction}
The China Space Station Telescope (CSST), the major science project of the China Manned Space Program, is a space telescope in the same orbit as the China Manned Space Station, which is about 400km height \citep{su2014two, gong2019cosmology}. It can dock with the space station for maintenance and upgrade. The basic information of CSST is shown in Tables~\ref{CSST_specifications}-\ref{CSST_mag}, respectively. With its wavelength coverage, high angular resolution, and large sky area coverage, CSST offers scientific opportunities and a great legacy value that complement other forthcoming space-based and ground-based surveys \citep{cao2018testing, zhan2018overview}. CSST is very suitable for astrometry studies of objects fainter than 20th magnitude. The CSST survey observations include multicolor imaging observations and seamless spectral observations. As suggested by the CSST Scientific Committee, CSST will focus on the following seven research directions: cosmology, galaxies and active galactic nucleus, the Milky Way and neighboring galaxies, stellar science, exoplanets and solar system objects, astrometry, transient sources and variable sources \citep{zhan2021wide}. Among them, astrometry, as a basic branch of astronomy, its main purpose is to provide necessary data for astronomy studies by measuring the position and motion of the celestial bodies, including planets and other solar system objects, stars in the Milky Way, galaxies and galaxy clusters in the universe. The astrometric results of CSST combine with its photometric information can provide the basic and reliable data for solving multiple core problems in the astronomical fields such as dark matter, dark energy, black holes, the Milky Way, and the solar system \citep{zhan2011consideration, gong2019cosmology, cao2022anisotropies}. Meanwhile, the CSST's astrometric observation data will complement Gaia's data, extending Gaia's tomographic mapping of the Milky Way to much fainter magnitudes, adding positions, parallaxes, distance estimates, color-magnitude-based stellar parameter estimates, and extinction estimates, which will bring great scientific opportunities and scientific wealth to CSST astronomical research. Furthermore, it will enable to reduce the Gaia precision degradation on positions and increase the number of available reference sources in the extragalactic regions (about 40\% of the sky) \citep{gai2022consolidation}.
 \begin{table}[htbp]
    \tiny
    \centering
   \caption[]{Key specifications of CSST. Table from \citet{zhan2021wide}.}
    \label{CSST_specifications}
    \begin{threeparttable}
   \begin{tabular}{ll}
     \toprule
     \toprule
      Items & Parameters\\
     \midrule
      Aperture & 2m \\
      Focal length & 28m \\
      Field of view & $\ge$ 1.1 $deg^2$ \\
      CCD plate scale of main focal plane & 0.073 arcsec/pixel\\
      Wavelength & 0.25$\sim$1.7$\mu$m, 590$\sim$730$\mu$m \\
      PSF $R_{EE80}$ $^a$ & $\le$ 0.15$''$ ($\lambda$=632.8nm, within field of view $<$ 1.1 $deg^2$) \\
      Pointing accuracy & LOS: $\le$ 5$''$, Roll: $\le$ 10$''$ \\
      Stability($<$ 300s) & LOS: $\le$ 0.05$''$, Roll: $\le$ 1.5$''$ (3$\sigma$) \\
      Jitter & $\le$ 0.01$''$ (3$\sigma$) \\
      Slew & 1$^\circ$/50s, 20$^\circ$/100s, 45$^\circ$/150s \\
      \bottomrule
    \end{tabular}%
    \begin{tablenotes} 
		\item$^{a}$ $R_{EE80}$ is the radius of 80\% energy concentration.
     \end{tablenotes} 
     \end{threeparttable}
\end{table}
\begin{table}[htbp]
    \tiny
   \centering
   \caption{Magnitude limit for multicolor imaging survey. Table from \citet{zhan2021wide}.}
     \begin{tabular}{ccccccccc}
     \toprule
     \toprule
     Sky area ($deg^2$) & Exposure time (s) & \multicolumn{7}{c}{Magnitude limit for different band$^a$} \\
     \midrule
           &       & \multicolumn{1}{c}{NUV} & \multicolumn{1}{c}{u} & \multicolumn{1}{c}{g} & \multicolumn{1}{c}{r} & \multicolumn{1}{c}{i} & \multicolumn{1}{c}{z} & \multicolumn{1}{c}{y}\\
 \cmidrule{3-9}    17500$^b$ & 150x2 & 25.4  & 25.4  & 26.3  & 26.0    & 25.9  & 25.2  & 24.4 \\
 \cmidrule{3-9}    400$^c$   & 250x8 & 26.7  & 26.7  & 27.5  & 27.2  & 27.0    & 26.4  & 25.7 \\
     \bottomrule
     \end{tabular}%
   \label{CSST_mag}%
       \begin{tablenotes} 
        \item$^{a}$ Point-source, 5$\sigma$, AB mag; 
		\item$^{b}$ Main sky survey areas; 
		\item$^{c}$ Selected deep fields.
     \end{tablenotes} 
 \end{table}%
\par
From Hipparcos \citep{perryman1997hipparcos} to Gaia \citep{prusti2016gaia}, the accuracy of position and parallax achieved a leap from 1 milli-arcsecond (mas) to a few tens of micro-arcseconds ($\mu$as) \citep{michalik2014joint}. For CSST, the solution accuracy of the five astrometric parameters (position, parallax, and proper motion) will be directly affected by the survey schedule, as will the output of other science applications that depend on observation cadence, directly or indirectly. A suitable survey strategy could be able to bring more observation samples and higher-precision observation data. The number and the distribution of the astrometric epoch data are the keys to obtaining the high-quality astrometric parameters of the celestial bodies. Therefore, how to effectively organize the survey schedule is very important to get the repeated observation data of the celestial bodies. At present, the scientific research team is designing a survey schedule plan that can meet the needs of multi-party observations. From the perspective of astrometry, the effectiveness of a survey strategy can be evaluated and optimized by analyzing the astrometric capability of CSST, and conversely, optimizing the survey strategy can also improve the astrometric capability of CSST. Improving the effectiveness of CSST's survey schedule is critical to improve CSST's scientific output. 

The CSST's astrometric capability simulation experiment is based on a self-designed astrometric solution software. Through simulation, we estimate the accuracy of the five astrometric parameters of stars at different magnitudes and different sky positions. The present paper gives a recipe for the practical realization of simulation evaluation of the astrometric capability of CSST. In Sect.~\ref{1jie}, we briefly describe the survey schedule used in this paper. In Sect.~\ref{2jie}, we establish the astrometric model based on the single-star kinematic model and the least-squares method to derive the astrometric parameters from simulated observation data. In Sect.~\ref{3jie}, we describe the steps for simulating observations based on a given survey schedule, which involves processing simulated observations using the astrometric model to obtain the astrometric solution for the observed celestial bodies. In Sect.~\ref{4jie}, we evaluate the astrometric capability of CSST by analyzing these astrometric solution. In Sect.~\ref{5jie}, we discuss the possibilities for improving the astrometric capability of CSST. Finally, we conclude this paper in Sect.~\ref{6jie}. 
\section{Survey strategy}\label{1jie}
The survey strategy is mainly based on operational scheduling constraints and observation requirements. The operational scheduling constraints include solar avoidance angle, lunar avoidance angle, Earth avoidance angle, field-of-view collocation, regular resupply, regular maintenance, regular orbit control, telescope maneuvering time and stabilization time, and so on. The observation requirements come from different scientific goals of CSST, for example, cosmological studies require CSST to observe in the median-to-high Galactic latitude and median-to-high Ecliptic latitude of the sky where the stellar density and zodiacal light background are low, and thus obtain a large sample of extragalactic objects \citep{zhan2021wide}.

\begin{figure}[htbp]
      
      \begin{minipage}{0.5\linewidth}
         \centering
         \includegraphics[scale=0.56]{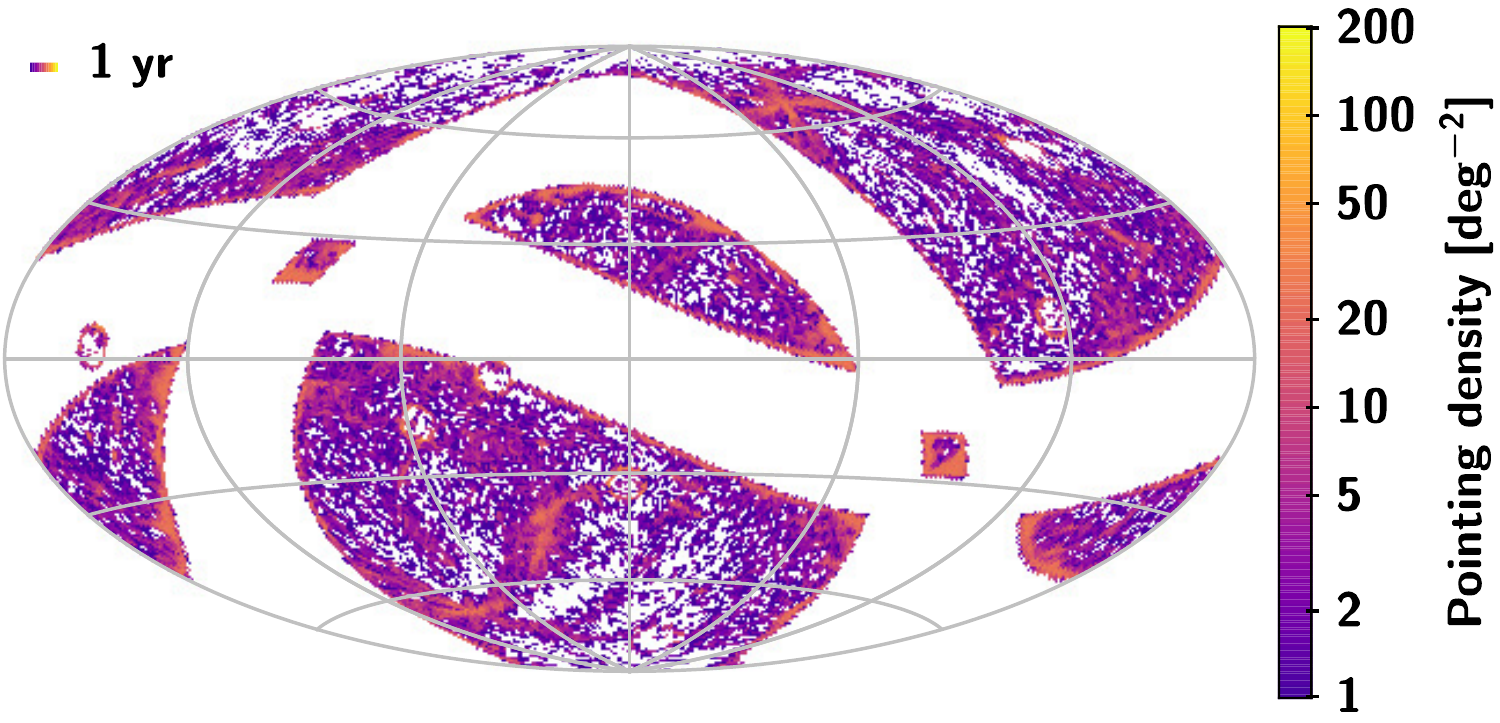}
      \end{minipage}%
      \begin{minipage}{0.5\linewidth}
         \centering
         \includegraphics[scale=0.56]{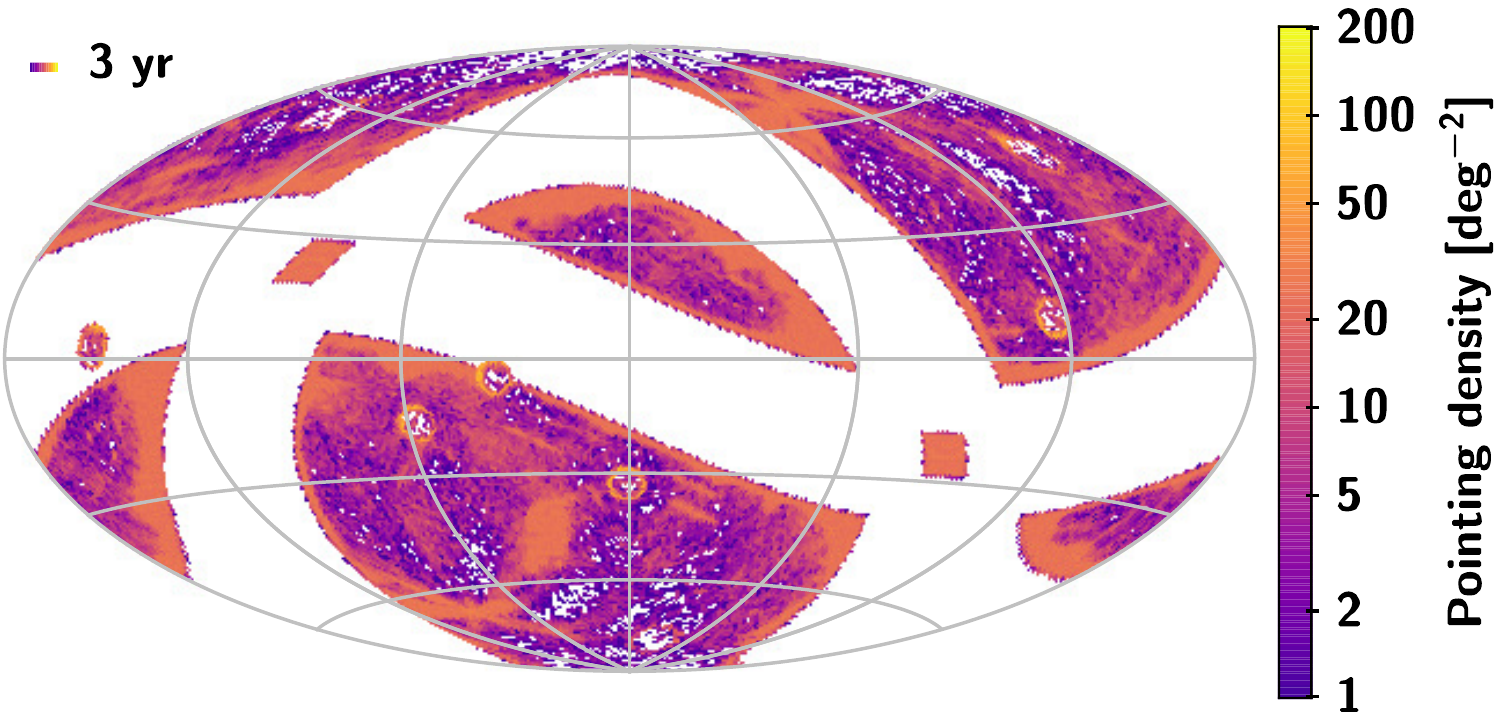}
      \end{minipage}
      
      \begin{minipage}{0.5\linewidth}
         \centering
         \includegraphics[scale=0.56]{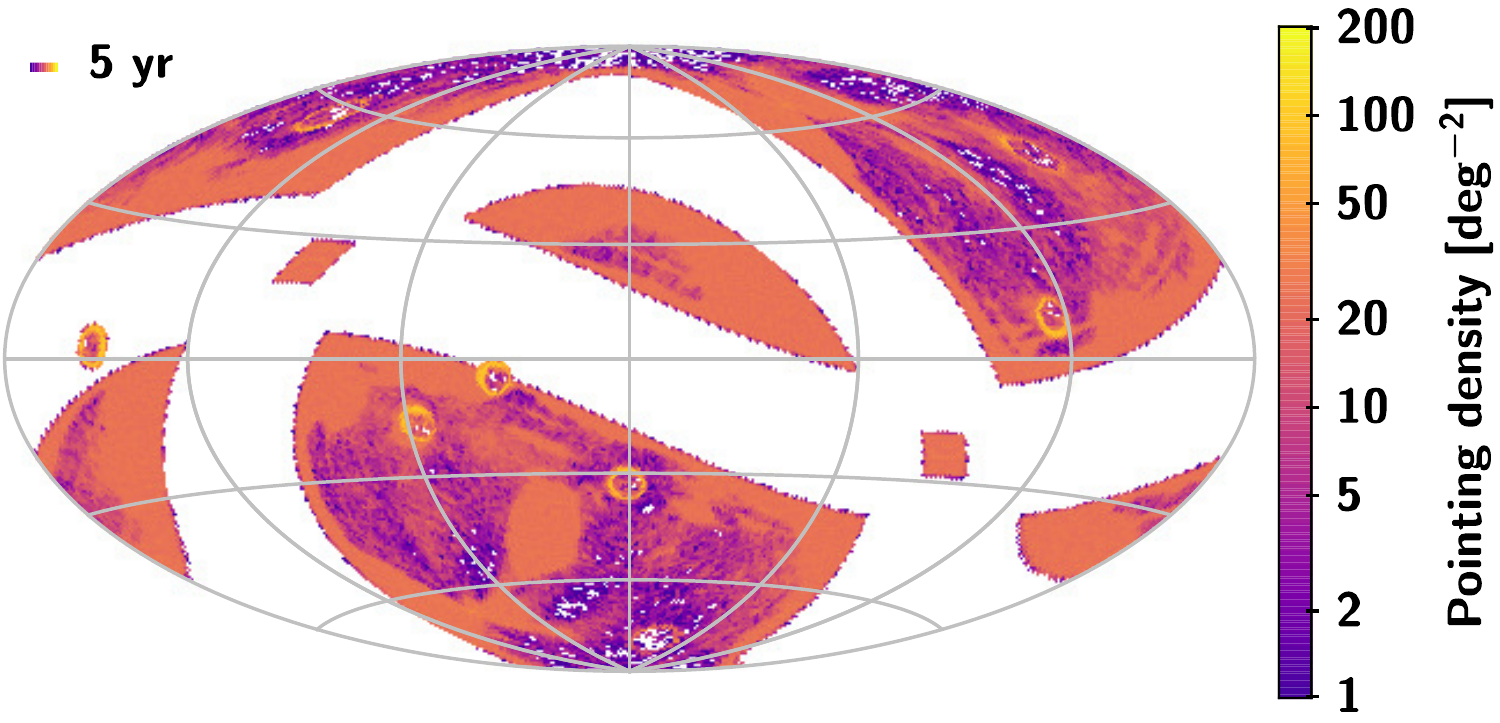}
      \end{minipage}%
      \begin{minipage}{0.5\linewidth}
         \centering
         \includegraphics[scale=0.56]{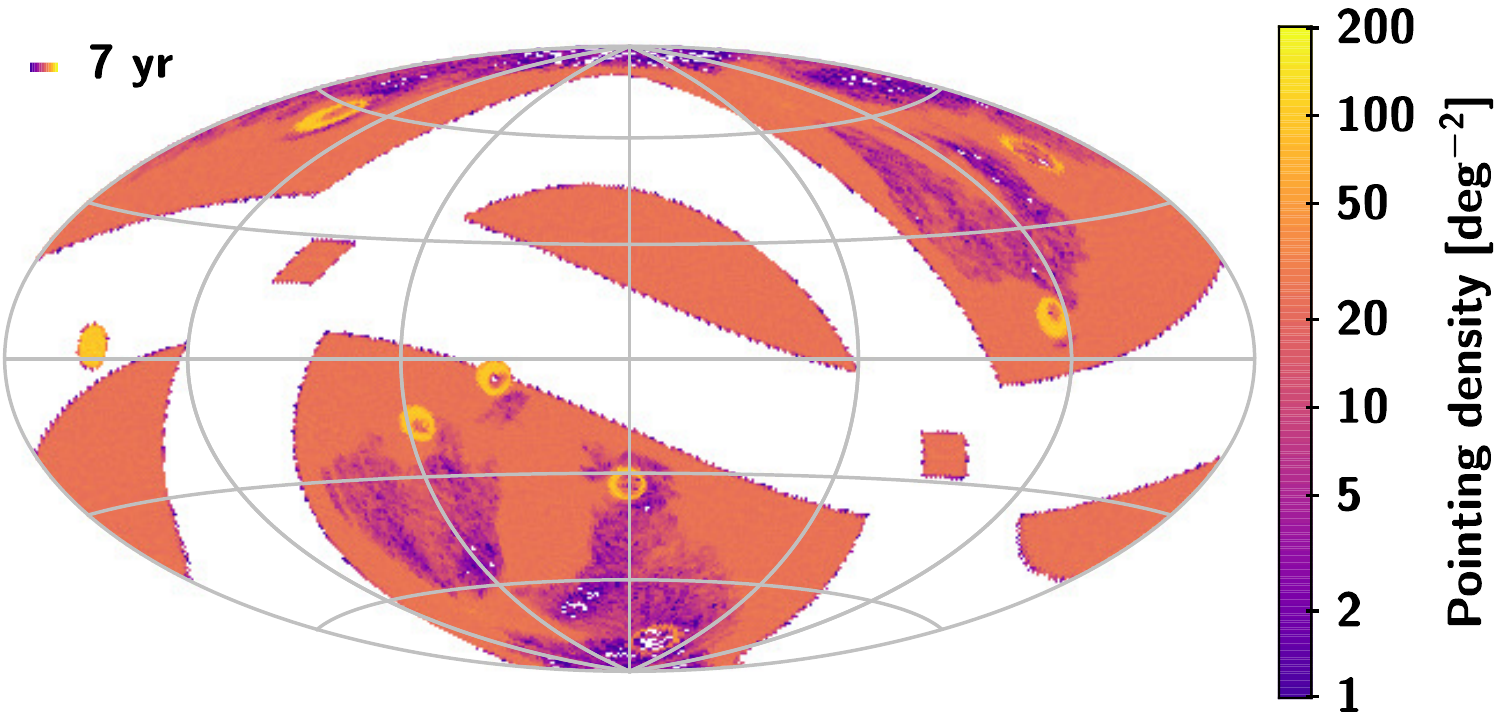}
      \end{minipage}
      
      \begin{minipage}{0.5\linewidth}
         \centering
         \includegraphics[scale=0.56]{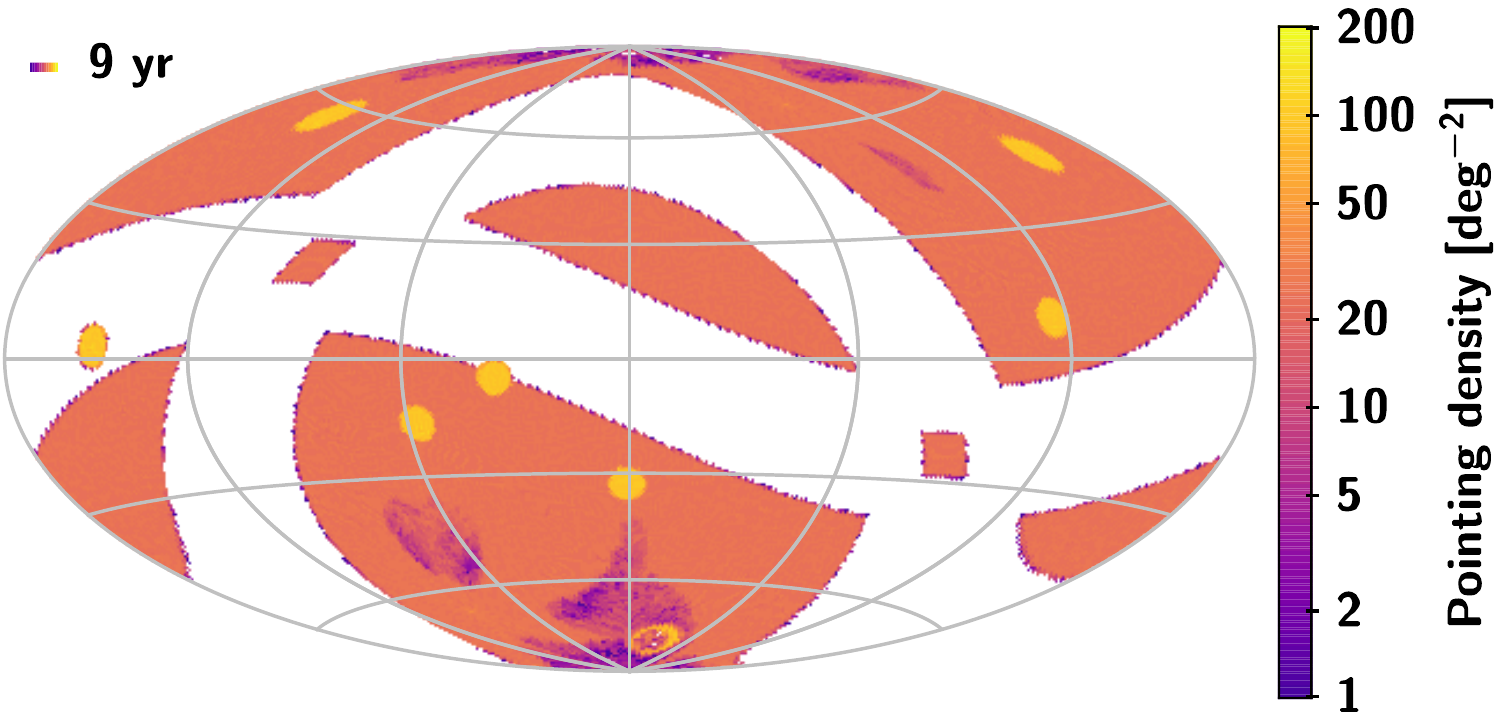}
      \end{minipage}%
      \begin{minipage}{0.5\linewidth}
         \centering
         \includegraphics[scale=0.56]{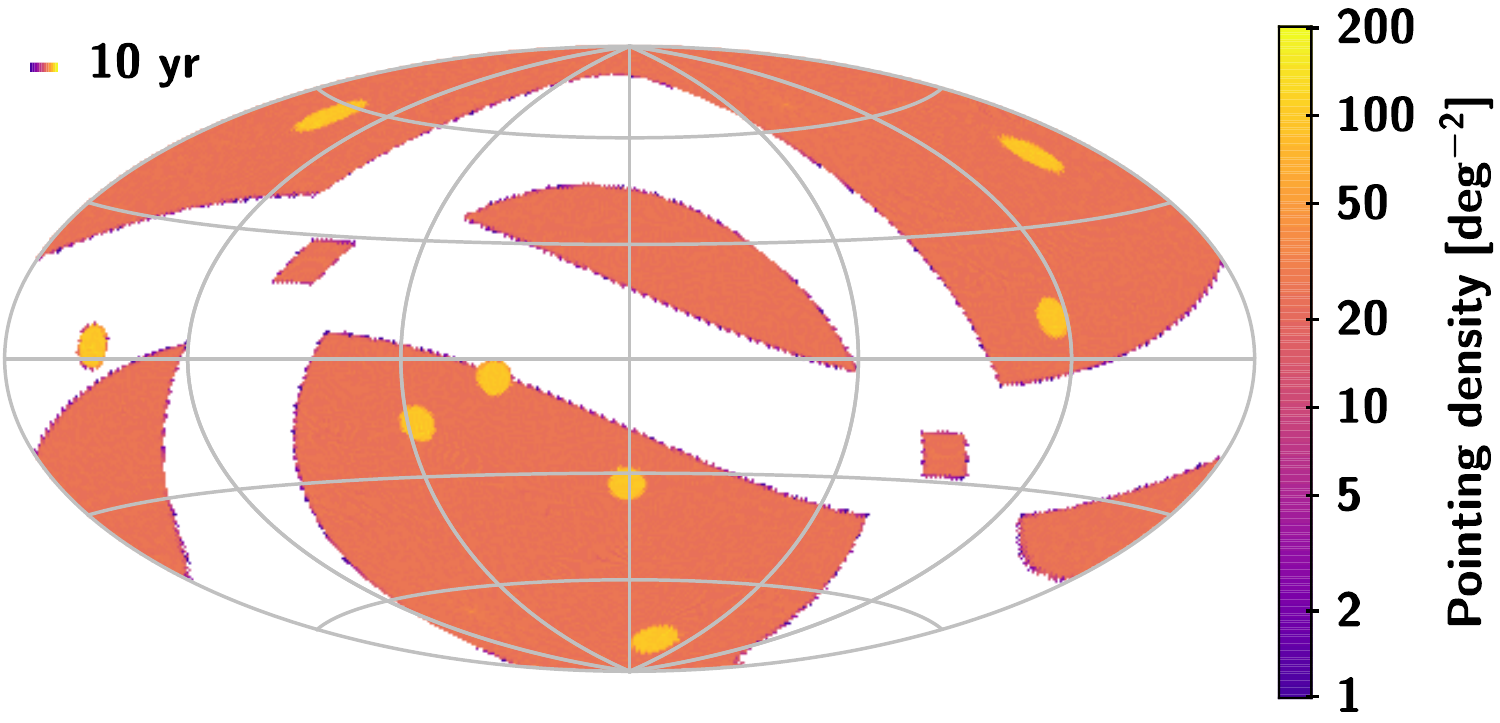}
      \end{minipage}
      
      \caption{Distribution of the observation pointing centers of the survey for 1, 3, 5, 7, 9 and 10 years, respectively. The yellow circles in the bottom right figure are the selected deep fields. All maps use an Aitoff projection in equatorial coordinates, with origin $\alpha=\delta=0$ at the centre and $\alpha$ increasing from right to left. Mean densities are shown for the observation pointing centers in cells of about 0.84 $deg^2$.}
       \label{csstxtbp}%
      \end{figure}

The CSST will allocate 70\% of its 10 years of operation time to image roughly 17500 square degrees of the sky. During the first year of the operation time, the image area will cover about 12000 square degrees of the sky\footnote{The area of 12000 square degrees is not covered by the entire band but by any band once.}, and images of NUV, u, g, r, i, z, and y bands (details can be found in Table~\ref{CSST_mag}) will be supplemented in each direction during the subsequent operation time \citep{zhan2021wide}. The observation pointing center distribution of the survey strategy\footnote{This is not the ultimate survey strategy, but only a typical case in the current study. The optimization of the survey strategy is still on going.} used in this paper is shown in Fig.~\ref{csstxtbp}. For a given survey schedule, the heterogeneity and concentration in the observation time directly affect the astrometric solution. Therefore, it is necessary to evaluate the CSST’s astrometric capability based on the given survey schedule and give independent suggestions for the optimization of the given survey schedule.
\section{Method}\label{2jie}
In this section, we will first introduce the single-star kinematic model, which is the basis for establishing the astrometric model and for simulating data on the variation of the object's position at different observation times. Then, we will introduce the astrometric model.
    \subsection{Single star motion}\label{2.1jie}
    Assume that the target celestial bodies we deal with are all single stars and such celestial bodies move linearly and uniformly relative to the solar system barycentre (SSB). For a single star whose proper motion in right ascension is $\mu_{\alpha \ast}$, proper motion in declination is $\mu_\delta$, and parallax is $\pi$, its motion (we ignore the radial motion) can be described by the following formula \citep[Vol. 1, Sect. 1.2.8]{perryman1997hipparcos}:
    \begin{equation}
      \label{DHXGCFC1}
      \mathbf{u}(t_\mathrm{i}) 
         = \langle \mathbf{r_{\mathrm{ep}}} + (t_\mathrm{i}-t_{\mathrm{ep}})
      (\mathbf{p_\mathrm{ep}} \mu_{\alpha \ast}
      +\mathbf{q_\mathrm{ep}} \mu_\delta)
      -\pi \frac{\mathbf{b_\mathrm{O}}(t_\mathrm{i})}{\mathrm{Au}} \rangle \,,
   \end{equation}
   where $ t_\mathrm{i}$ is the time of observation; $ t_{\mathrm{ep}}$ is the reference epoch; $\mathbf{b_\mathrm{O}}(t_\mathrm{i})$ is the barycentric position of CSST at the time of observation; $\mathrm{Au}$ is the astronomical unit; $\mathbf{u}(t_\mathrm{i})$ is the unit vector of  the coordinate direction of the star observed by CSST: 
   \begin{equation}
     \label{GCFCLBJ1}
     \mathbf{u}(t_\mathrm{i}) = 
        \begin{bmatrix} \cos \delta(t_\mathrm{i}) \cdot \cos \alpha(t_\mathrm{i})\\
        \cos \delta(t_\mathrm{i}) \cdot \sin \alpha(t_\mathrm{i})\\
        \sin \delta(t_\mathrm{i})
        \end{bmatrix} \,;
   \end{equation}
  $\alpha(t_\mathrm{i})$ and $\delta(t_\mathrm{i})$ are the observed values of the right ascension and declination of the star, respectively; $\mathbf{r_\mathrm{ep}}$ is the unit vector along the line between the star and SSB at the reference epoch; $\mathbf{p_\mathrm{ep}}$ and $\mathbf{q_\mathrm{ep}}$ are the unit vectors in the directions of increasing $\alpha_\mathrm{ep}$ and $\delta_\mathrm{ep}$ at $\mathbf{r_\mathrm{ep}}$, respectively: 
  \begin{eqnarray}
      \mathbf{r_{\mathrm{ep}}} = 
       \begin{bmatrix} 
       \cos \delta_\mathrm{ep} \cos \alpha_\mathrm{ep}\\
       \cos \delta_\mathrm{ep} \sin \alpha_\mathrm{ep}\\
       \sin \delta_\mathrm{ep}
       \end{bmatrix} ,\ 
       \mathbf{p_\mathrm{ep}} = 
       \begin{bmatrix}
       -\sin \alpha_\mathrm{ep}\\
          \cos \alpha_\mathrm{ep}\\
          0
       \end{bmatrix} ,\ 
       \mathbf{q_\mathrm{ep}} =
       \begin{bmatrix} 
       -\sin \delta_\mathrm{ep} \cos \alpha_\mathrm{ep}\\
       -\sin \delta_\mathrm{ep} \sin \alpha_\mathrm{ep}\\
       \cos \delta_\mathrm{ep}
       \end{bmatrix}  \,;
    \end{eqnarray}
    $\alpha_\mathrm{ep}$ and $\delta_\mathrm{ep}$ are the right ascension and declination of the star at the reference epoch, respectively; $\langle \rangle$ denotes vector normalisation: $\langle \mathbf{a}\rangle=\frac{\mathbf{a}}{|\mathbf{a}|}$.
    \subsection{Astrometric model of the single star}\label{2.2jie}
To facilitate the astrometric solution of the single star, we use the Local Plane Coordinate (LPC) to describe the motion of the single star. The base vectors for the LPC are $\mathbf{r_{\mathrm{ep}}}$, $\mathbf{p_{\mathrm{ep}}}$ and $\mathbf{q_{\mathrm{ep}}}$. At the time $t_\mathrm{i}$, the position component of the single star in the LPC along the base vectors $\mathbf{p_{\mathrm{ep}}}$ and $\mathbf{q_{\mathrm{ep}}}$ directions are $\xi$ and $\eta$, respectively\citep[Vol. 1, Sect. 1.2.9]{perryman1997hipparcos}:
    \begin{equation}
      \label{LPCgs1}
      \xi(t_\mathrm{i}) = \frac{\mu_{\alpha^{\ast}}(t_\mathrm{i}-t_\mathrm{ep})-\mathbf{p_\mathrm{ep}^\prime} \mathbf{b_\mathrm{O}}(t_\mathrm{i})\pi /Au}{1-\mathbf{r_\mathrm{ep}^\prime}\mathbf{b_\mathrm{O}}(t_\mathrm{i})\pi /Au} + \Delta \alpha^{\ast} \,,
   \end{equation}
   \begin{equation}
      \label{LPCgs2}
      \eta(t_\mathrm{i}) = \frac{\mu_\delta(t_\mathrm{i}-t_\mathrm{ep})-\mathbf{q_\mathrm{ep}^\prime}\mathbf{b_\mathrm{O}}(t_\mathrm{i})\pi /Au}{1-\mathbf{r_\mathrm{ep}^\prime}\mathbf{b_\mathrm{O}}(t_\mathrm{i})\pi /Au} + \Delta \delta \,,
   \end{equation}
   where the prime ($^\prime$) denotes scalar product; $\Delta \alpha^{\ast}=\Delta\alpha\cos\delta_\mathrm{ep}$ and $\Delta \delta$ are the offsets at the reference epoch in $\alpha_\mathrm{ep}$ and $\delta_\mathrm{ep}$, respectively.
   \par
We refer to Eqs.~(\ref{LPCgs1},\ref{LPCgs2}) as the single observation equations. In the equations, the unknown parameters are $\Delta \alpha^{\ast}$, $\Delta \delta$, $\pi$, $\mu_{\alpha^\ast}$ and $\mu_\delta$, which are also the astrometric parameters that we want to solve for. In Appendix \ref{1app}, we describe how to solve these astrometric parameters using the least squares method. We also estimate the standard uncertainty of the five astrometric parameters with the method of precision estimation of adjustment of indirect observations. This model, which can solve five astrometric parameters at the same time, is marked as the astrometric 5-parameter solution model. We do not estimate parallax for targets that cannot solve for effective parallax due to insufficient number of observations. We refer it as 4-parameter solution.
\section{Simulations}\label{3jie}
   As described by \citet{michalik2014joint}, the simulations are carried out in the following 3 steps: 1) Create a catalogue of all the stars used to generate CSST observations, namely the input catalogue. The input catalogue is also used to evaluate the solvability of the astrometric parameters. 2) Simulate the observations of the stars based on the given survey schedule (see Fig.~\ref{csstxtbp}) and the magnitude dependence of the astrometric uncertainty. 3) Analysis the solvability of the astrometric parameters and generate the final catalogue.
   \par
Through the statistical analysis of the astrometric solution in the final catalogue, the astrometric capability of CSST can be evaluated. In this section, we only cover the process of simulating the observation and processing the observation data, while the detailed evaluation of the astrometric capability of CSST is described in Sect.~\ref{4jie}.
    \subsection{Input catalogue}
\begin{figure}[htbp]
       
      \begin{minipage}{\linewidth}
         \centering
         \includegraphics[scale=0.56]{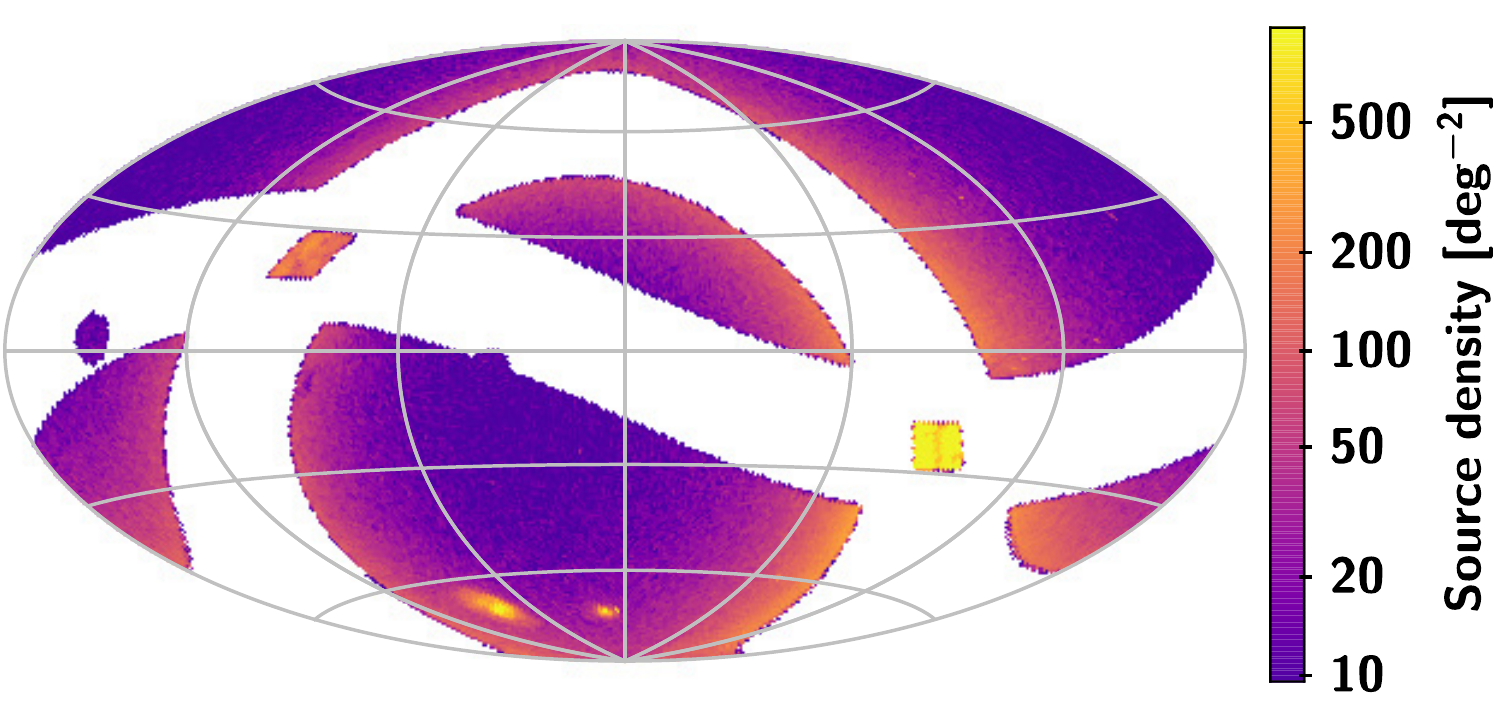}
      \end{minipage}
      
       \caption{Distribution of the sources in the input catalogue, about $1256640$ sources. The map uses an Aitoff projection in equatorial coordinates, with origin $\alpha=\delta=0$ at the centre and $\alpha$ increasing from right to left. Mean density is shown for the sources in cells of about 0.84 $deg^2$.}
      \label{srxbfb}%
      \end{figure}

\begin{figure}[htbp]
       
      \begin{minipage}{\linewidth}
         \centering
         \includegraphics[scale=0.56]{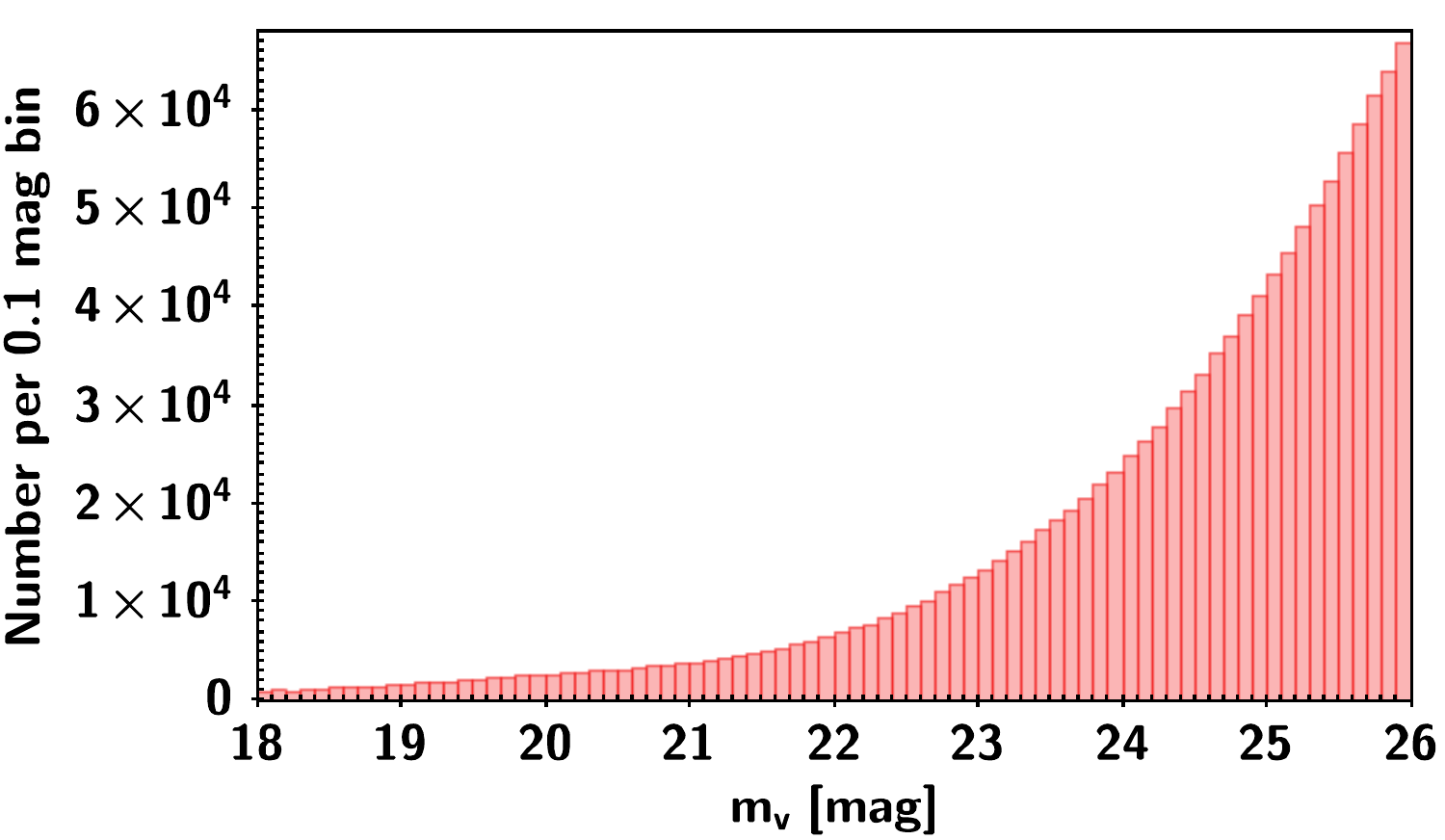}
      \end{minipage}
      
       \caption{Distribution of the magnitudes for the input catalogue shown as histograms with bins of 0.1 mag in width.}
      \label{srxbsmxdfb}%
      \end{figure}

To make the celestial bodies in the input catalogue as realistic as possible, we randomly extract about 1.2 million celestial bodies from the Gaia DR3 \citep{brown2021gaia, lindegren2021gaia, vallenari2022gaia}\footnote{The Gaia DR3 catalogue is the outcome of analyzing raw data from the first 34 months of the Gaia mission. Description at \url{https://cosmos.esa.int/web/gaia/dr3}; data available on \url{https://gea.esac.esa.int/archive/}} and assign a random magnitude between 18 and 26 mag to each target but keep the astrometric parameters from Gaia. The magnitudes are assigned with reference to the distribution of the number of sources in different magnitude intervals\footnote{Data available on \url{https://gea.esac.esa.int/archive/visualization/}} in the Gaia DR3, and by extrapolating this distribution we obtain a distribution of the number of sources in the magnitude interval 18-26 mag\footnote{CSST plans to survey the sky in NUV, u, g, r, i, z and y bands, however, for simplicity in this simulation, we concentrate in g band, which has a magnitude limit from around 18 to 26mag.}. The input catalogue contains information about the ``true'' position, proper motion, parallax and magnitude of these celestial bodies. The density of the sources and distribution of the magnitudes of the sources in the input catalogue are shown in Fig.~\ref{srxbfb} and Fig.~\ref{srxbsmxdfb}, respectively.
    \subsection{Simulating CSST observations}
Simulated observations are produced by combining the input catalogue with the survey schedule with the following two main steps:
    \begin{enumerate}
       \item The information of the survey schedule includes: all observed times, the barycentre coordinates of CSST and the directions of the observation pointing centers at the time of observation. The celestial bodies from the input catalogue observed at each observed times are calculated. For these observed celestial bodies, the corresponding positions of the coordinate directions at each observed times can be solved by Eq.~(\ref{DHXGCFC1}).
       \item The observation errors are simulated as Gaussian noise based on the magnitude dependence of the astrometric uncertainty provided by the CSST astrometry team. The magnitude distribution of the astrometric uncertainty is shown in Fig.~\ref{mngcwc}. The astrometric uncertainty takes into account various physical and instrumental effects, such as cosmic ray, non-linearity, distortion, sky background, dark current, bias, flat field, charge diffusion effect, CCD saturation overflow, failed image elements/columns, instrument platform jitter, gain, readout noise, etc. The difference between the astrometric uncertainty of CSST and the theoretical reference lower bound $\sigma_{lb}$ is because the former takes into account many errors, especially instrumental effects, such as distortion, etc. To simulate the observed star positions more realistically, we use the magnitude dependence of the astrometric uncertainty. For the observations of a celestial body at magnitude $g$, the observation errors at the right ascension and declination can be treated as Gaussian noise with the normal distribution $N(0,\sigma_{\alpha^{\ast}}^2)$ and $N(0,\sigma_{\delta}^2)$, respectively. 
           \par
          \begin{figure*}[htbp]
            \centering
            \includegraphics[scale=0.45]{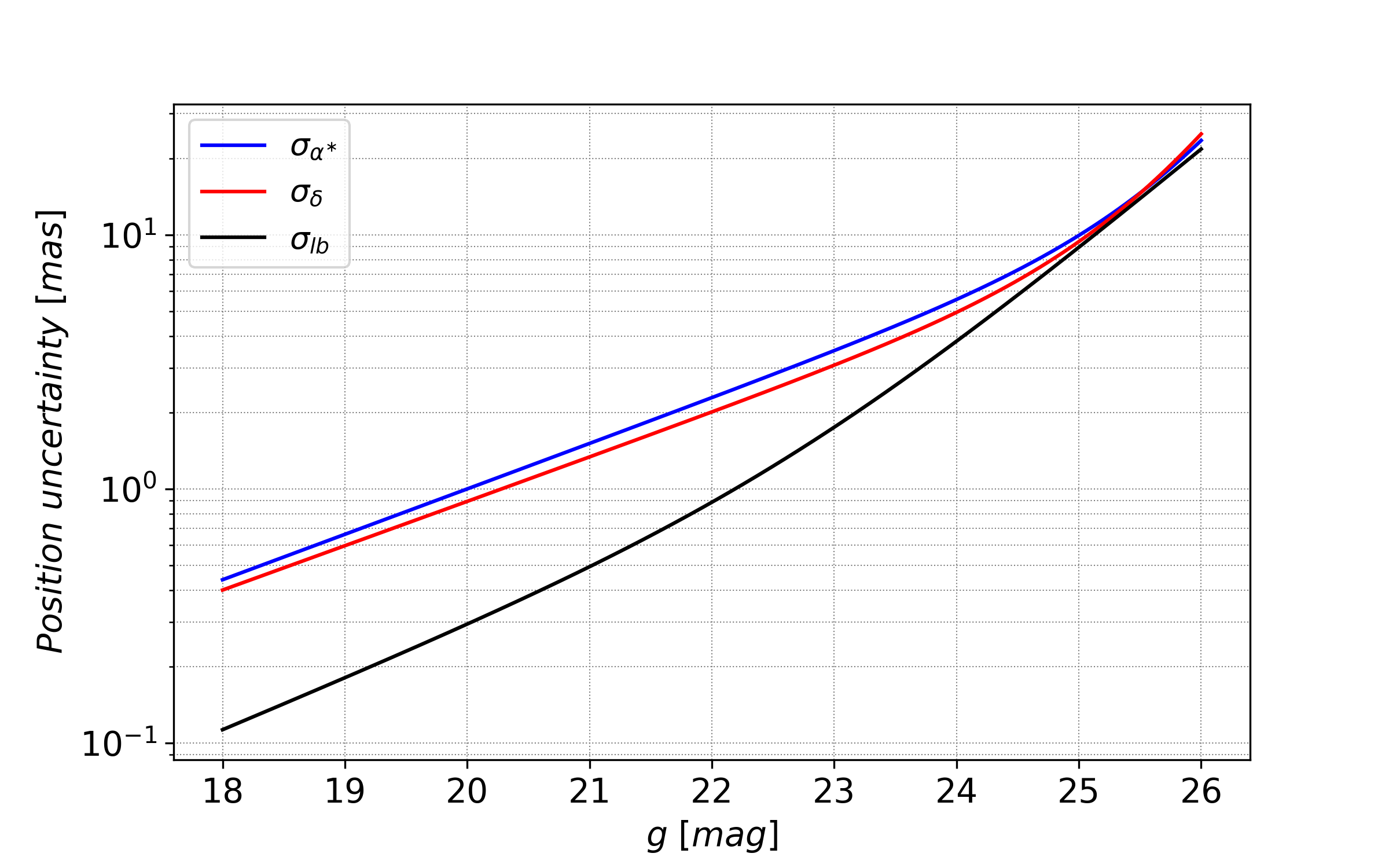}
            \caption{Magnitude dependence of the astrometric uncertainty in $\alpha$ and $\delta$. The blue solid line is the standard uncertainty of the right ascension, the red solid line is the standard uncertainty of the declination, and the black solid line is the theoretical reference lower bound of the positional precision $\sigma_{lb}$ for a diffraction-limited image (see Appendix \ref{2app}).}
             \label{mngcwc}%
            \end{figure*}
    \end{enumerate}
    \subsection{Final catalogue}
    The astrometric model in Sect.~\ref{2.2jie} is used to process simulated observation data. Then we compare the solutions with the ``true'' parameter values in the input catalogue. The selected astrometric parameters compatible with the true values within three times the respective uncertainty are considered as effective parameters, also known as solvable parameters. The relevant information (solved value, standard uncertainty, correlation coefficient, residual, and so on) of the effective parameters is used to form the final catalogue.
\section{Results}\label{4jie}
    \subsection{Overview}
\begin{figure*}[htbp]
      \centering
      \includegraphics[scale=0.65]{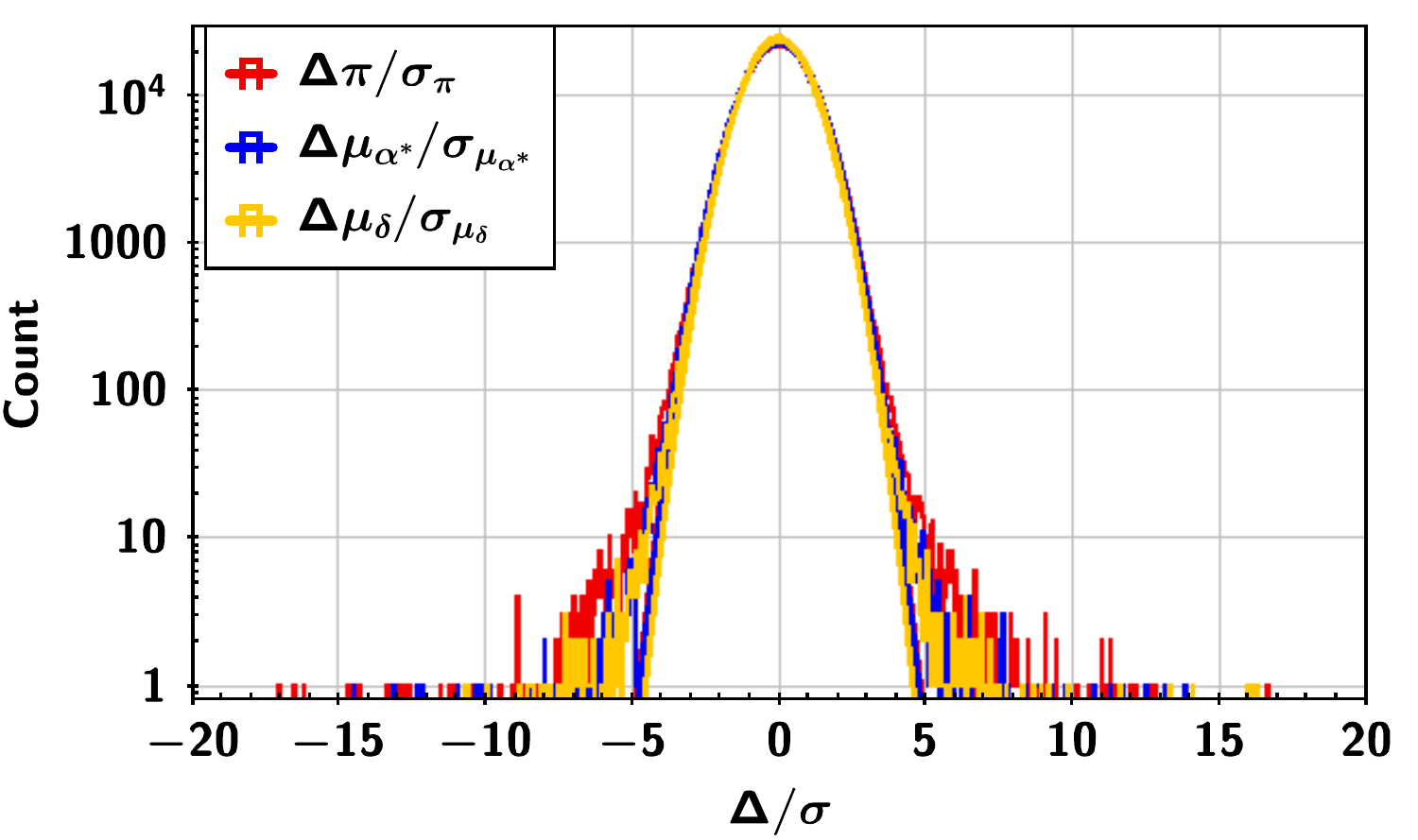}
      \caption{Distribution of the difference ($\Delta$ in the figure) between the solutions and the true values for the parallax and proper motion. The three curves show the corresponding best-fit Gaussian distributions.}
               \label{Para_err}%
   \end{figure*}

    The final catalogue contains the astrometric solution for 1211330 sources (96.39\% of the total number of sources in the input catalogue), among which the 5-parameter solvable sources accounted for 97.18\% (1177160 sources) of the total number of sources in the final catalogue, and the 4-parameter (position offsets and proper motion) solvable sources accounted for 98.57\% (1194059 sources) (details can be found in Table~\ref{number_s}). The distribution of the difference between the solutions and their respective true values can be found in Fig.~\ref{Para_err}. The corresponding standard deviations of the best-fit Gaussian distributions for parallax and proper motion are 1.073, 1.061 and 1.014, respectively.
     \begin{table}[htbp]
     \tiny
     \centering
   \caption[]{Number of the effective parameters for all sources in the final catalogue.}
    \label{number_s}
   \begin{tabular}{lr}
     \toprule
     \toprule
      Parameter type & Number of sources (percentage) \\
     \midrule
      Total & 1211330 (100\%) \\
      5-parameter & 1177160 (97.18\%) \\
      4-parameter$^a$ & 1194059 (98.57\%) \\
      Right ascension & 1204495 (99.44\%) \\
      Declination & 1205863 (99.55\%) \\
      Parallax & 1193698 (98.54\%) \\
      Proper motion in right ascension & 1204271 (99.42\%) \\
      Proper motion in declination & 1205743 (99.54\%) \\
      \bottomrule
    \end{tabular}%
           \begin{tablenotes} 
		\item$^{a}$ The 5-parameter sources are included.
     \end{tablenotes} 

    \end{table}
    \subsection{Predicted astrometric capability of CSST}
Next, we will evaluate the astrometric capability of CSST by analyzing the following information from the final catalogue:
    \begin{itemize}
      \item the standard uncertainties of the astrometric parameters: $\sigma_{\alpha^{\ast}}$, $\sigma_{\delta}$, $\sigma_\pi$, $\sigma_{\mu_{\alpha^{\ast}}}$ and $\sigma_{\mu_\delta}$; 
      \item the ten correlation coefficients among the five parameters: $\rho_{\alpha^{\ast} \delta}$, $\rho_{\alpha^{\ast} \pi}$, $\rho_{\alpha^{\ast} \mu_{\alpha^{\ast}}}$, $\rho_{\alpha^{\ast} \mu_\delta}$, $\rho_{\delta \pi}$, $\rho_{\delta \mu_{\alpha^{\ast}}}$, $\rho_{\delta \mu_\delta}$, $\rho_{\pi \mu_{\alpha^{\ast}}}$, $\rho_{\pi \mu_\delta}$ and $\rho_{\mu_{\alpha^{\ast}} \mu_\delta}$; 
      \item the signal-to-noise ratio of parallax: $SNR_{\pi}\ (SNR_{\pi}=\frac{\pi}{\sigma_{\pi}})$; 
      \item the median epoch of celestial observation time series: median epoch\footnote{We mark the beginning time as zero, the end time is the 10Th year.};
      \item  the mean epoch of celestial observation time series: mean epoch;
      \item the standard deviation of celestial observation time series (which describes the dispersion of observation series of each source): $\sigma_t$; and the mean of $\sigma_t$ of each source: $\sigma_{mean}$.
      
    \end{itemize}
    \par
\begin{figure*}[htbp]
        \begin{minipage}{0.5\linewidth}
            \centering
              \includegraphics[scale=0.56]{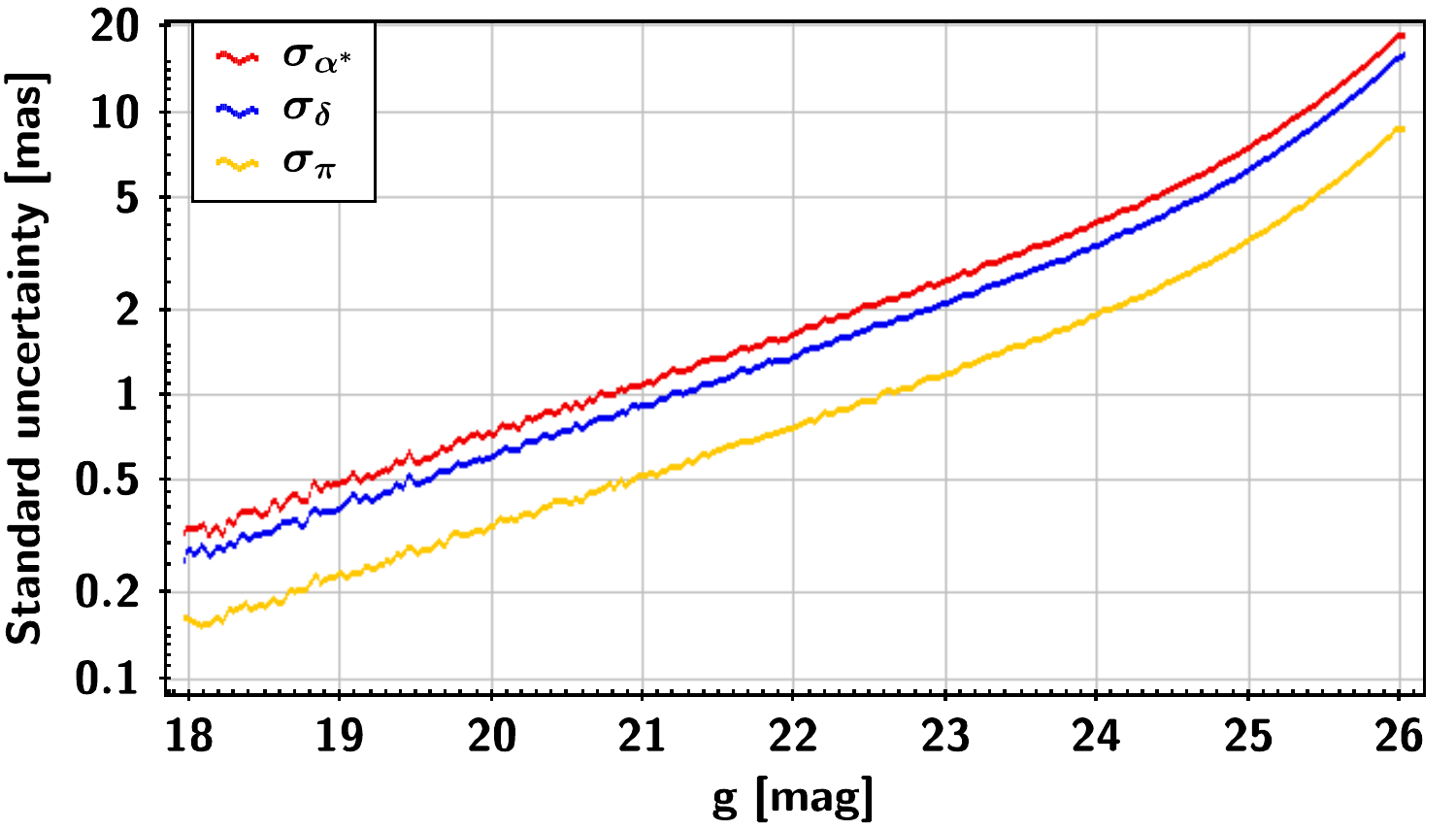}   
         \end{minipage}%
         \begin{minipage}{0.5\linewidth}
               \centering \includegraphics[scale=0.56]{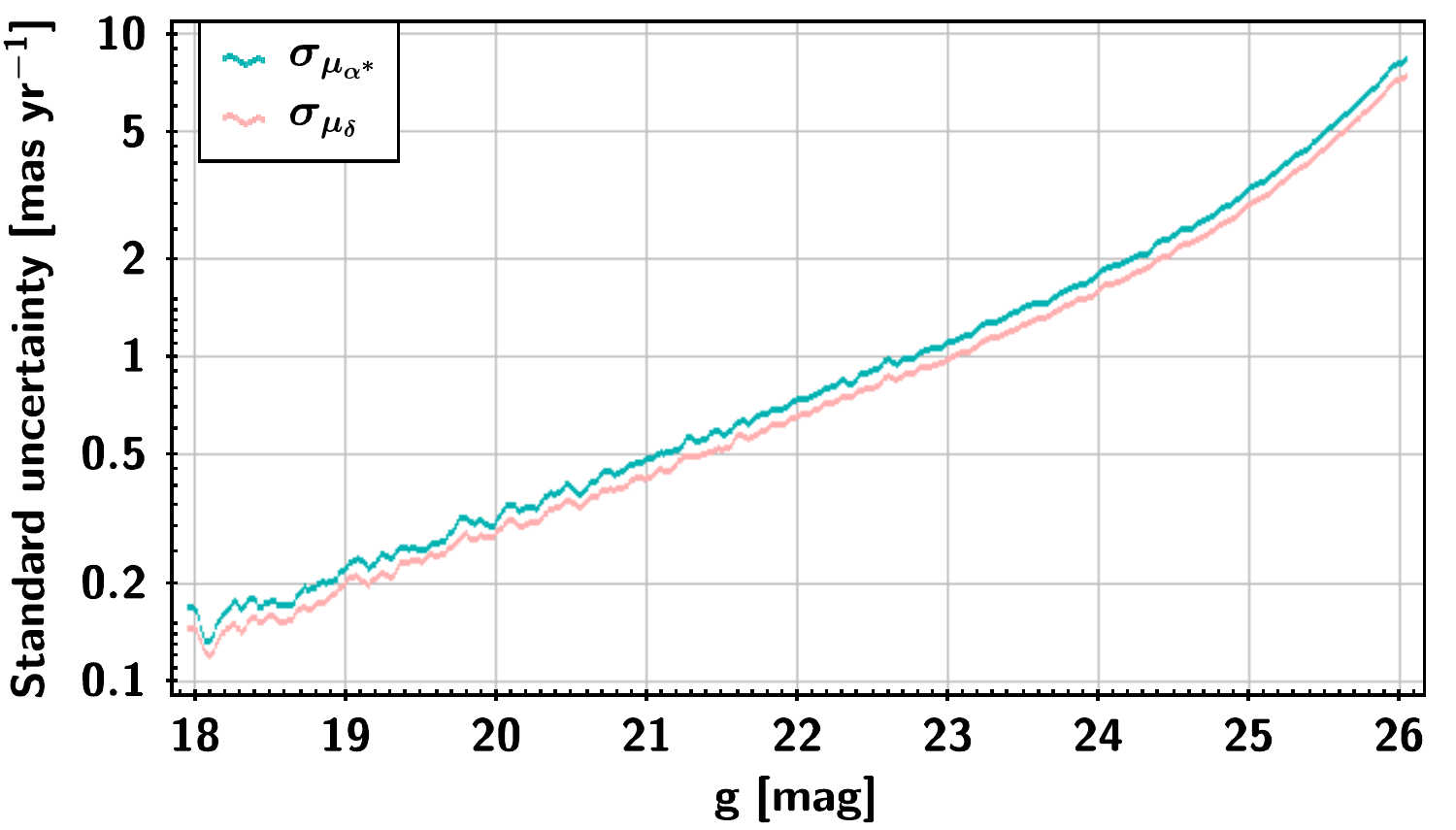}   
         \end{minipage}%
         
      \caption{Median standard uncertainties of the astrometric parameters for different magnitudes.}
      \label{Vbaifenshu}%
    \end{figure*}
\begin{landscape}
      \begin{figure}[!h]
      
         \begin{minipage}{0.2\linewidth}
               \centering
               \includegraphics[scale=0.3]{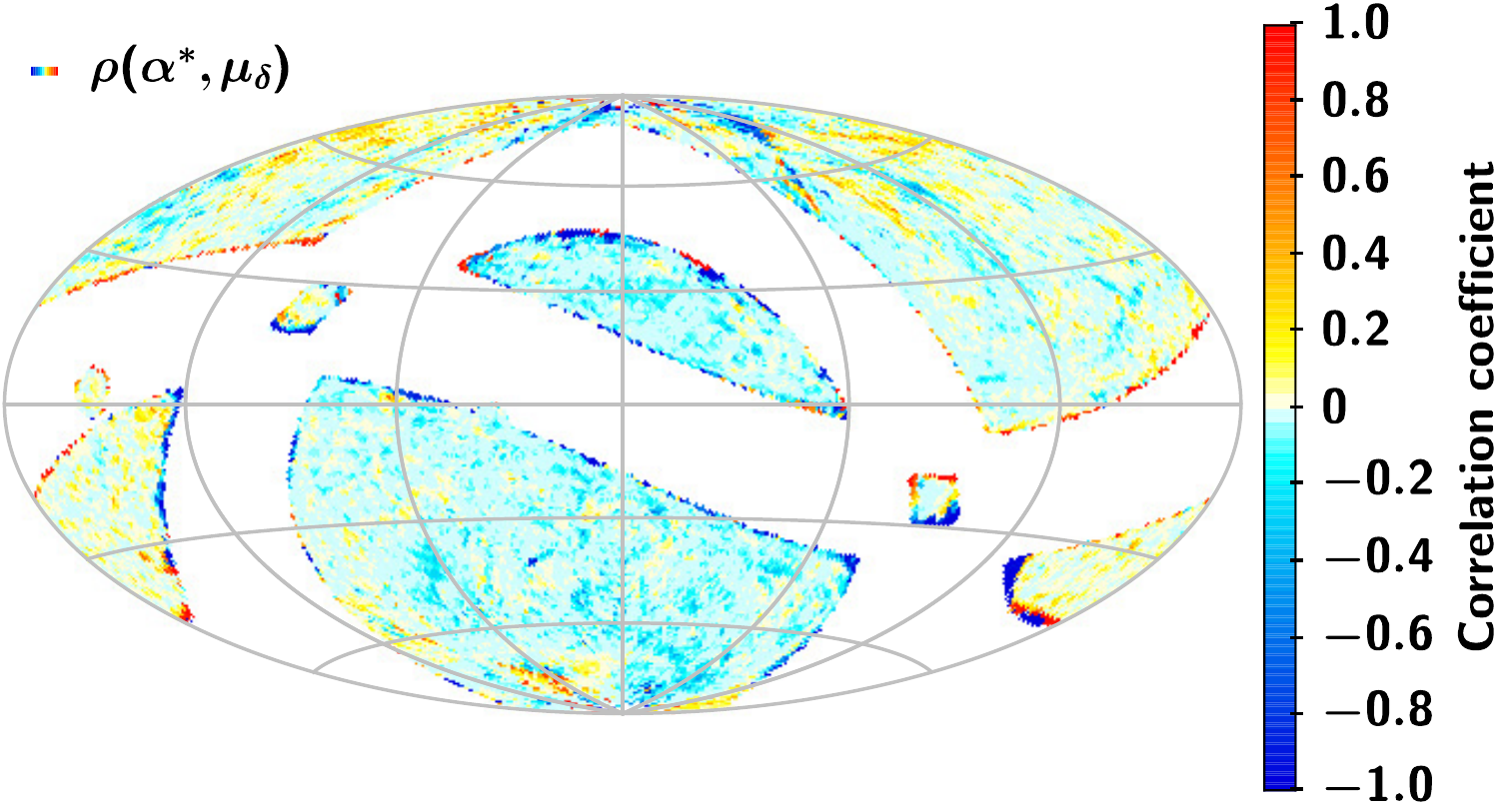}
         \end{minipage}%
         \begin{minipage}{0.2\linewidth}
              \centering
              \includegraphics[scale=0.3]{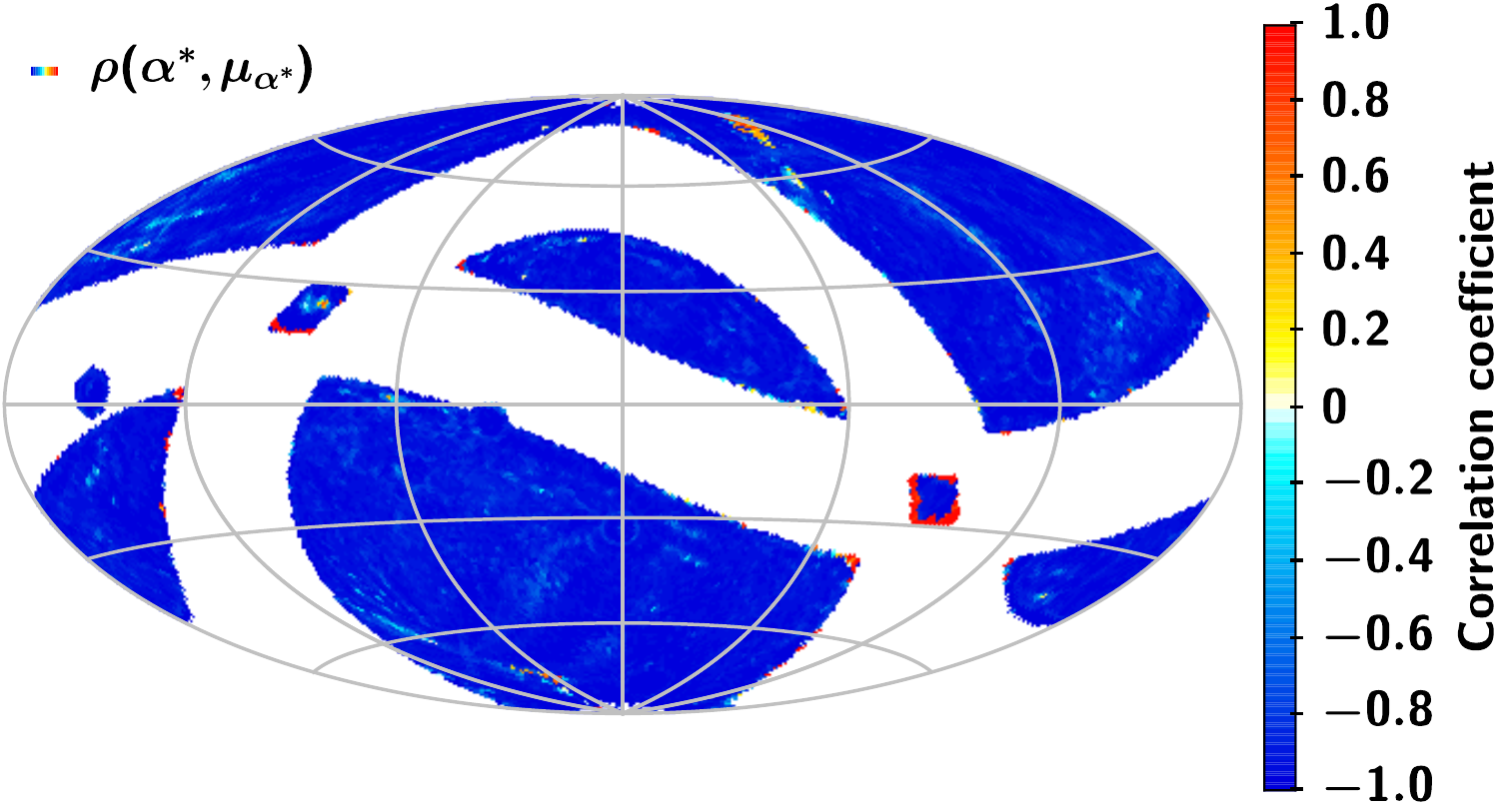}   
         \end{minipage}%
         \begin{minipage}{0.2\linewidth}
               \centering
               \includegraphics[scale=0.3]{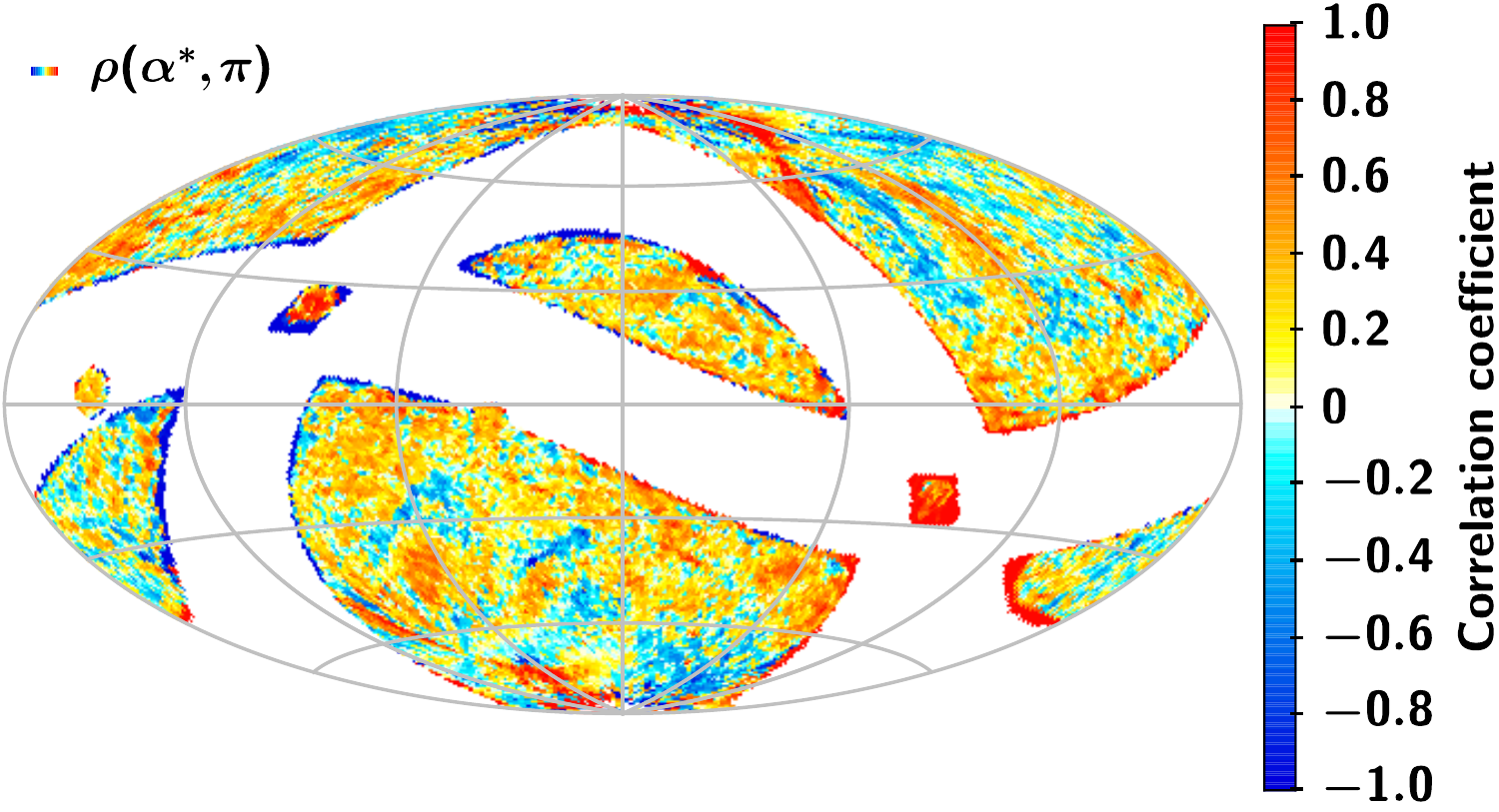}   
         \end{minipage}%
         \begin{minipage}{0.2\linewidth}
               \centering
               \includegraphics[scale=0.3]{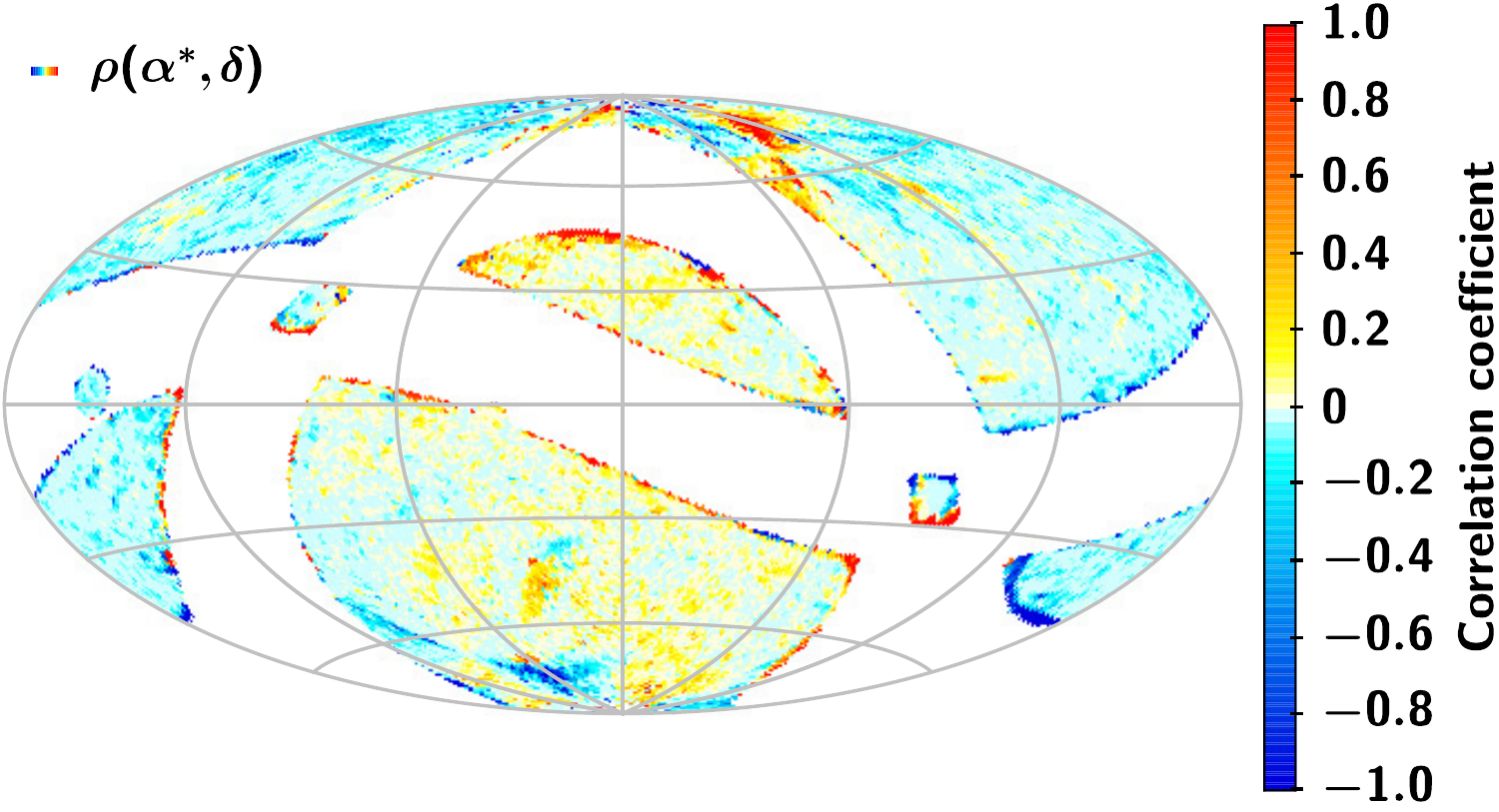}   
         \end{minipage}%
         \begin{minipage}{0.2\linewidth}
               \centering
               \includegraphics[scale=0.3]{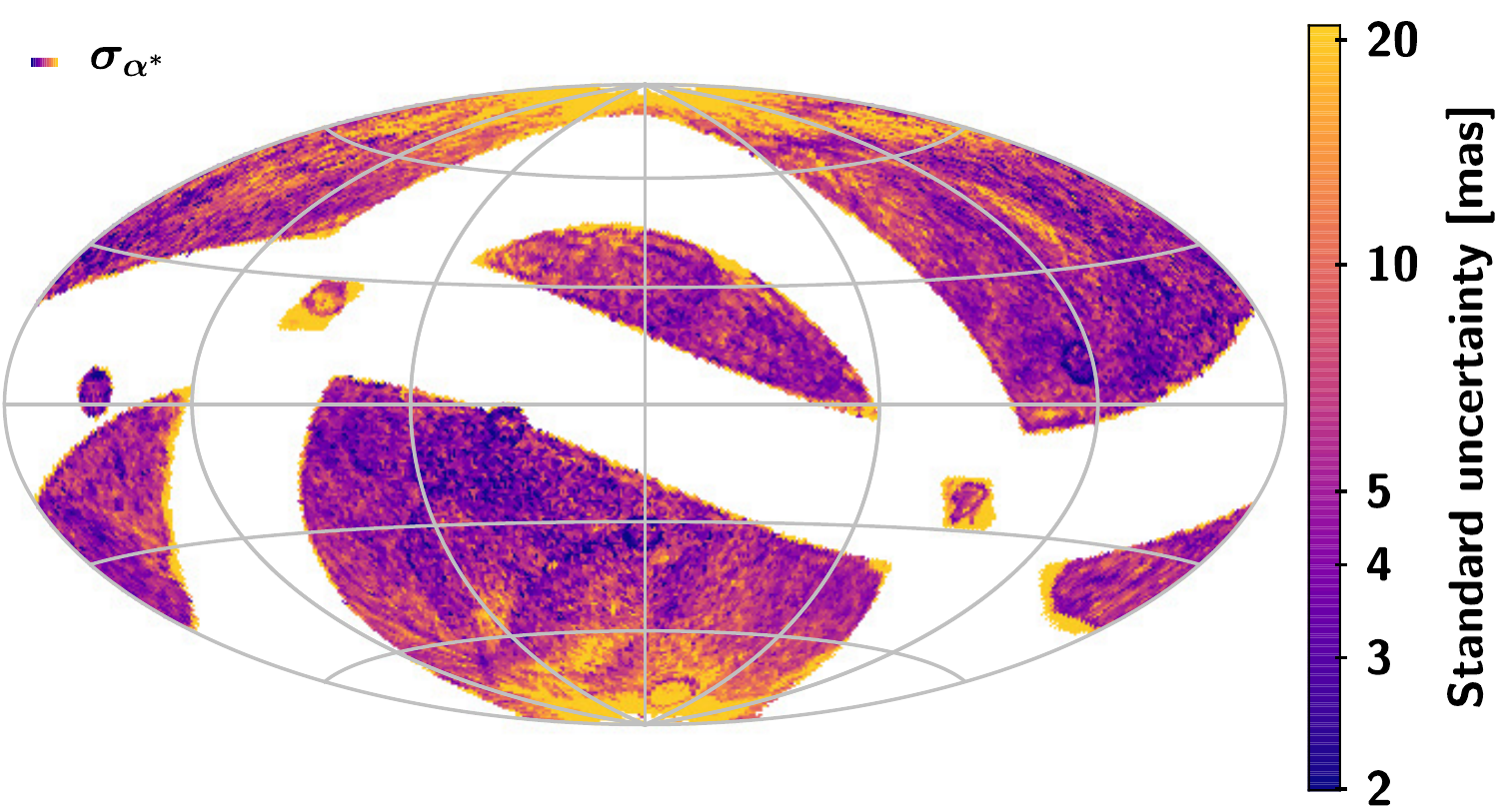}   
         \end{minipage}
         \\
         \\
         \\
         \begin{minipage}{0.2\linewidth}
               \centering
               \includegraphics[scale=0.3]{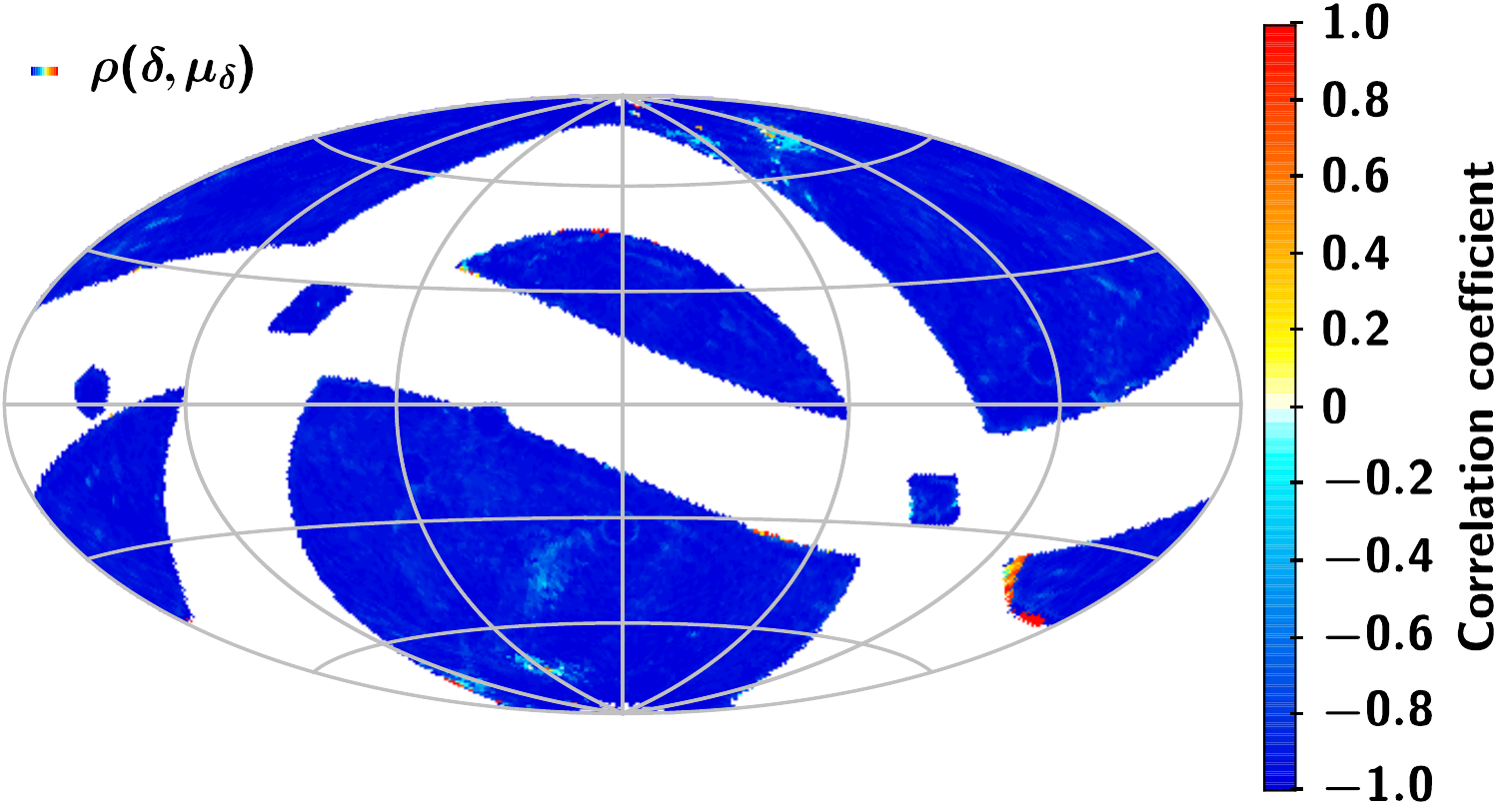}   
         \end{minipage}%
         \begin{minipage}{0.2\linewidth}
               \centering
               \includegraphics[scale=0.3]{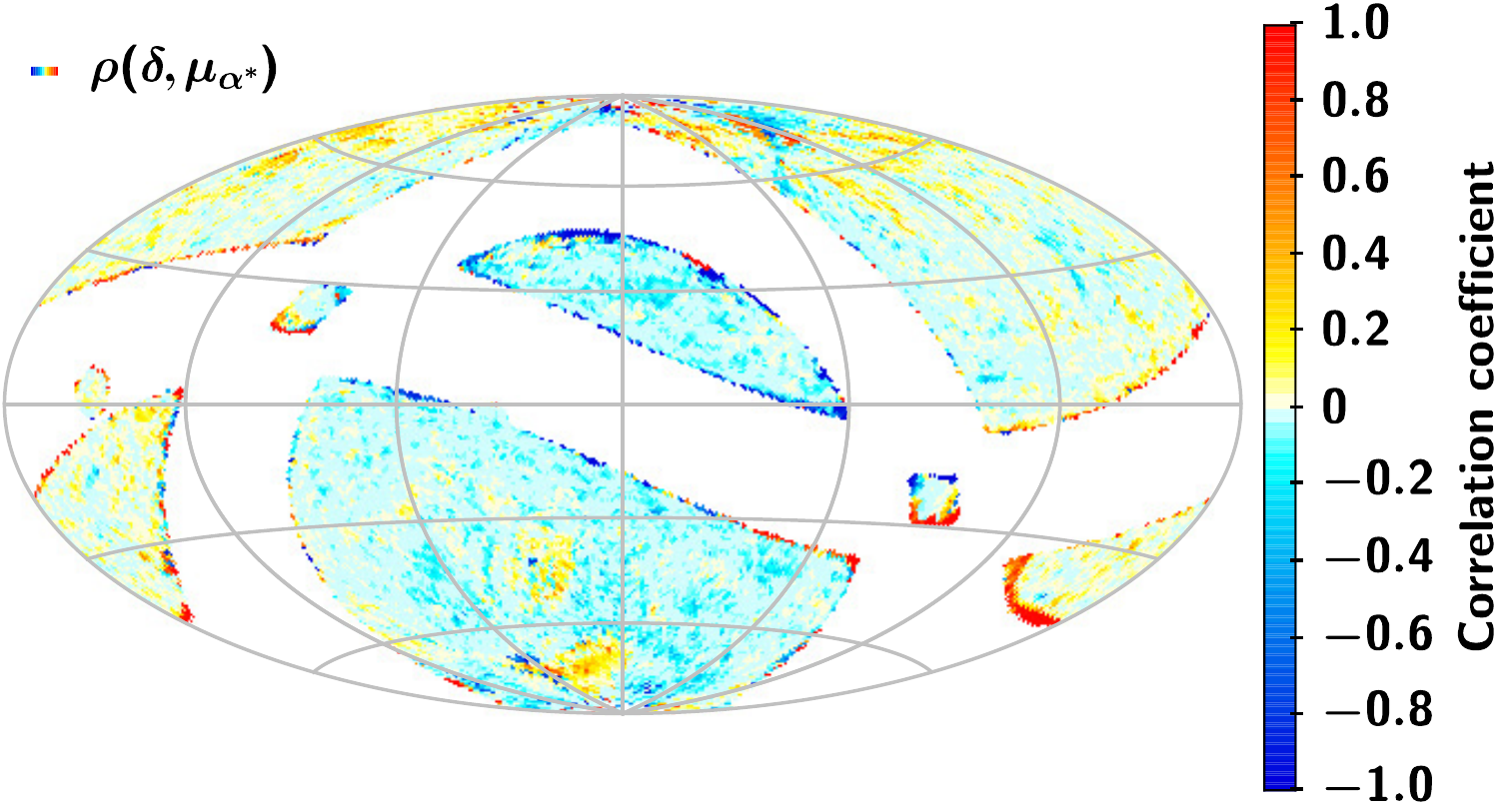}   
         \end{minipage}%
         \begin{minipage}{0.2\linewidth}
               \centering
               \includegraphics[scale=0.3]{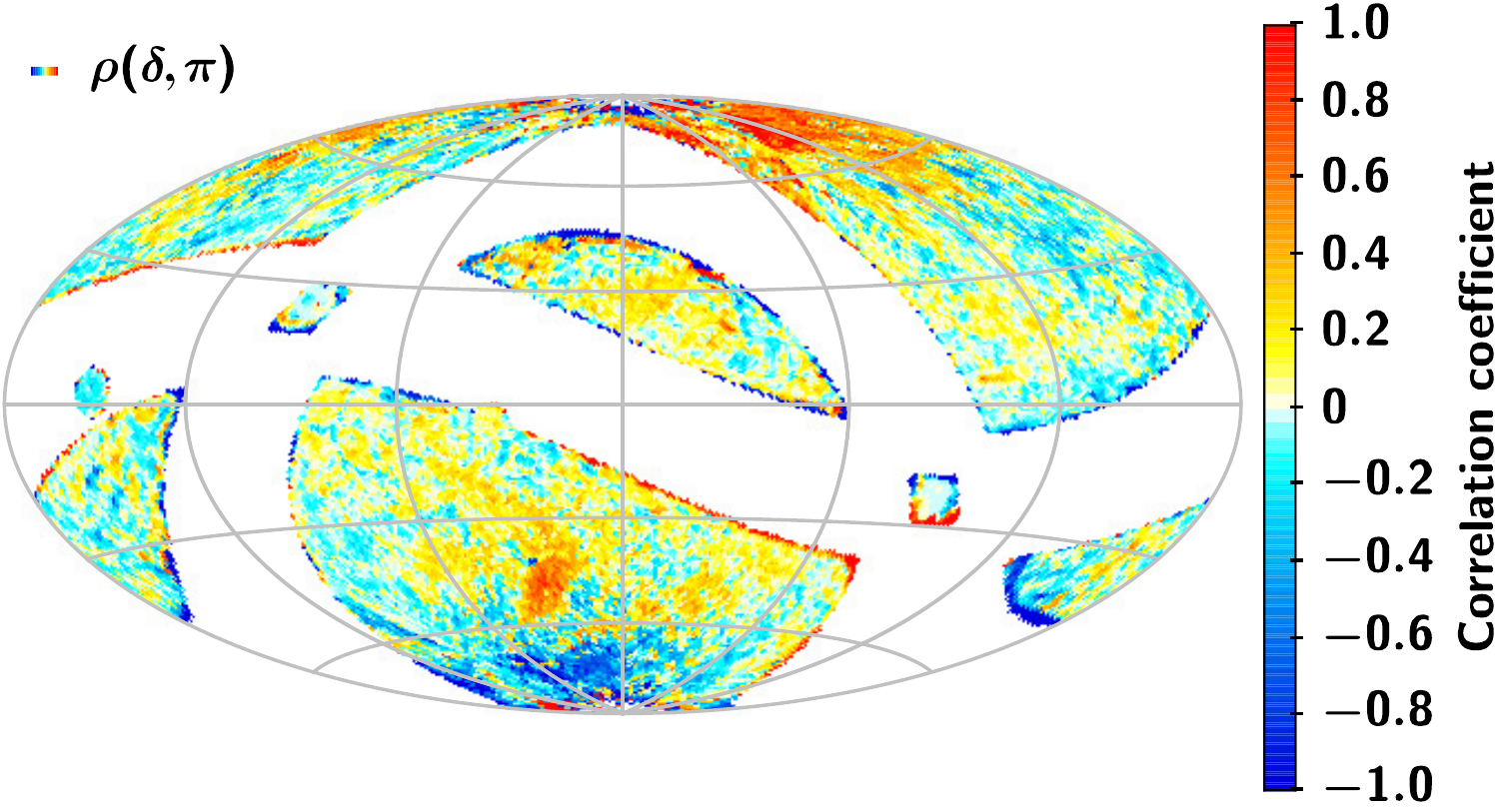}   
         \end{minipage}%
         \begin{minipage}{0.2\linewidth}
               \centering
               \includegraphics[scale=0.3]{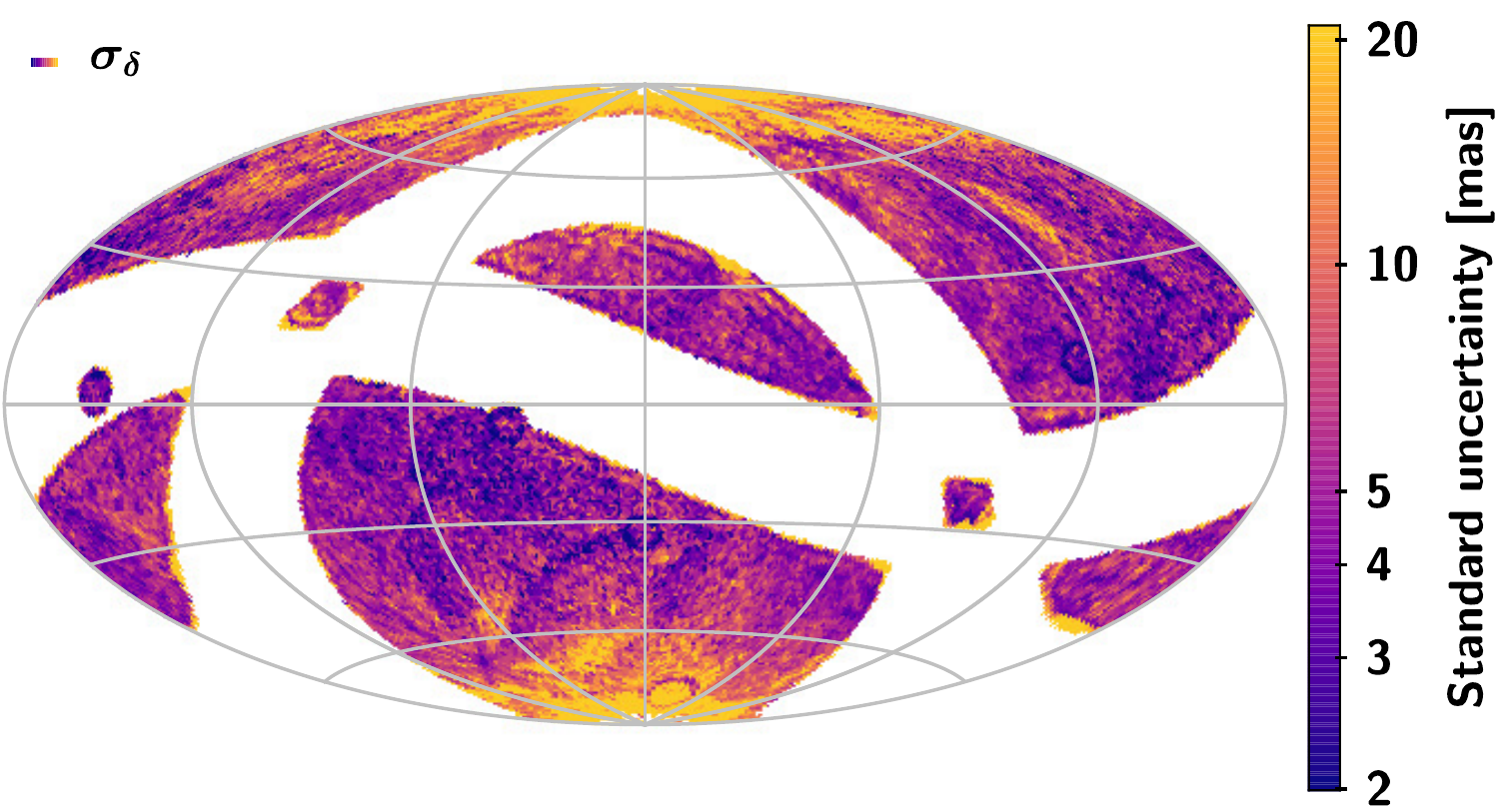}   
         \end{minipage}
         \\
         \\
         \\
         \begin{minipage}{0.2\linewidth}
               \centering
               \includegraphics[scale=0.3]{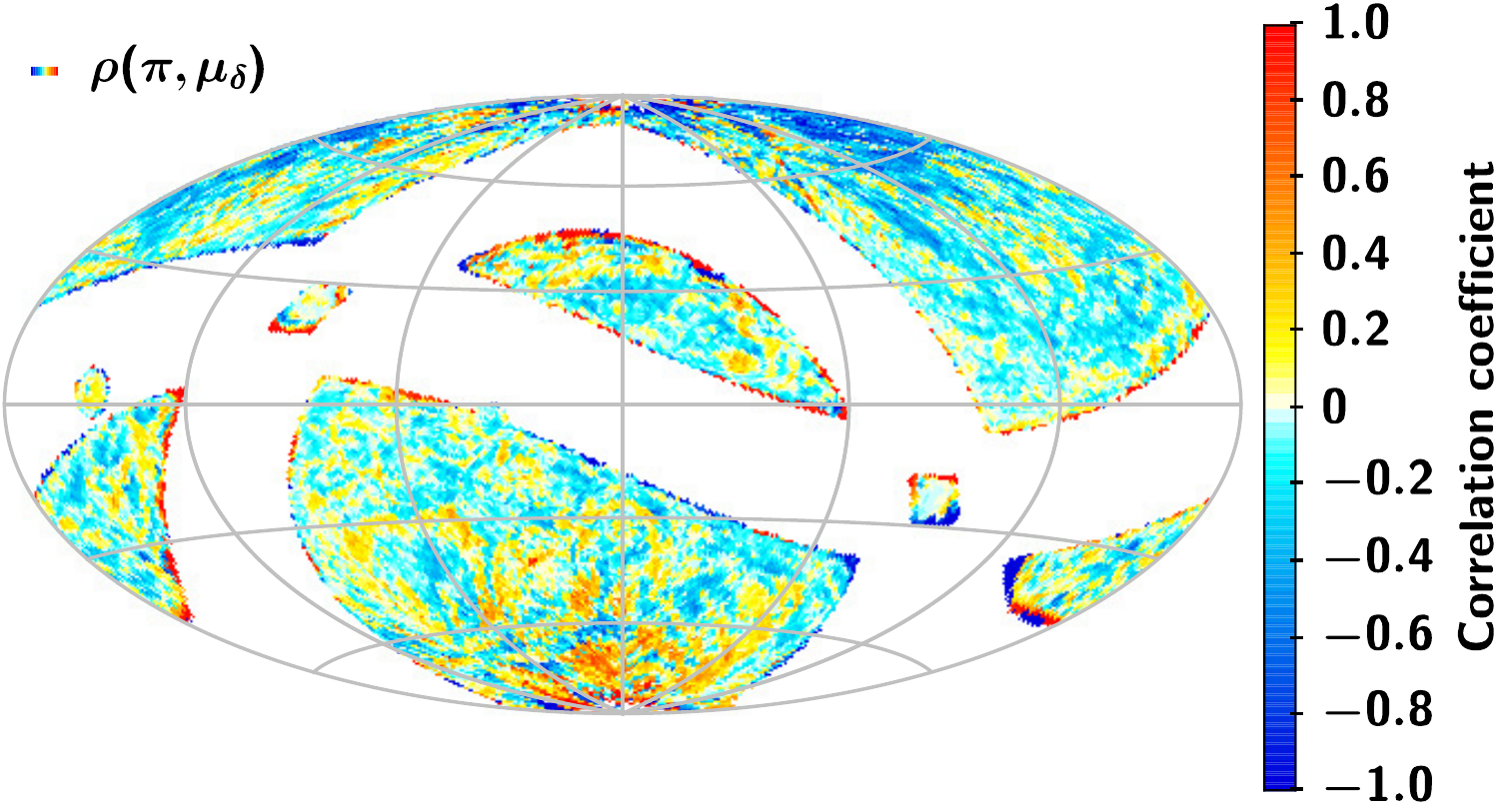}   
         \end{minipage}%
         \begin{minipage}{0.2\linewidth}
               \centering
               \includegraphics[scale=0.3]{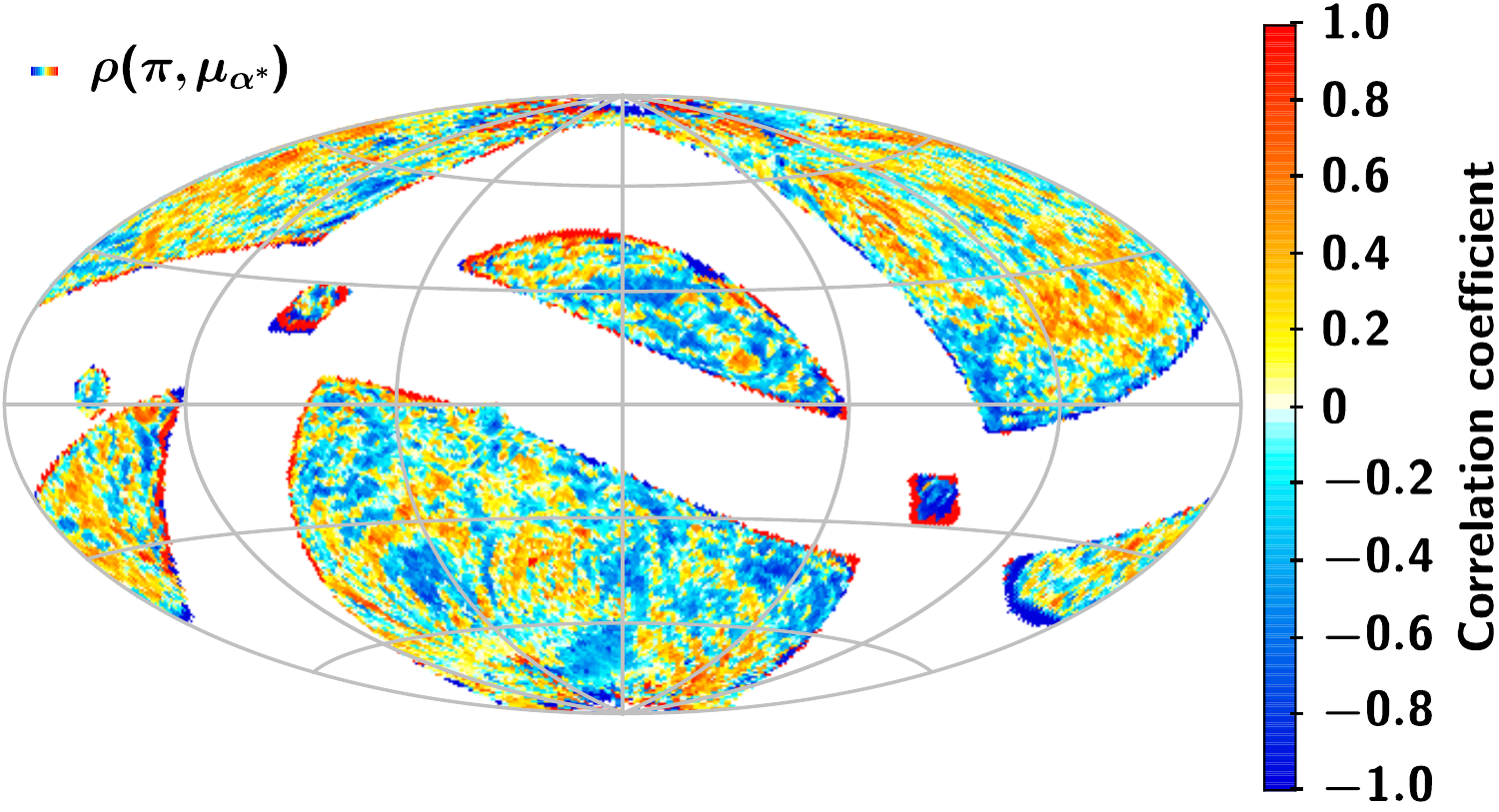}   
         \end{minipage}%
         \begin{minipage}{0.2\linewidth}
               \centering
               \includegraphics[scale=0.3]{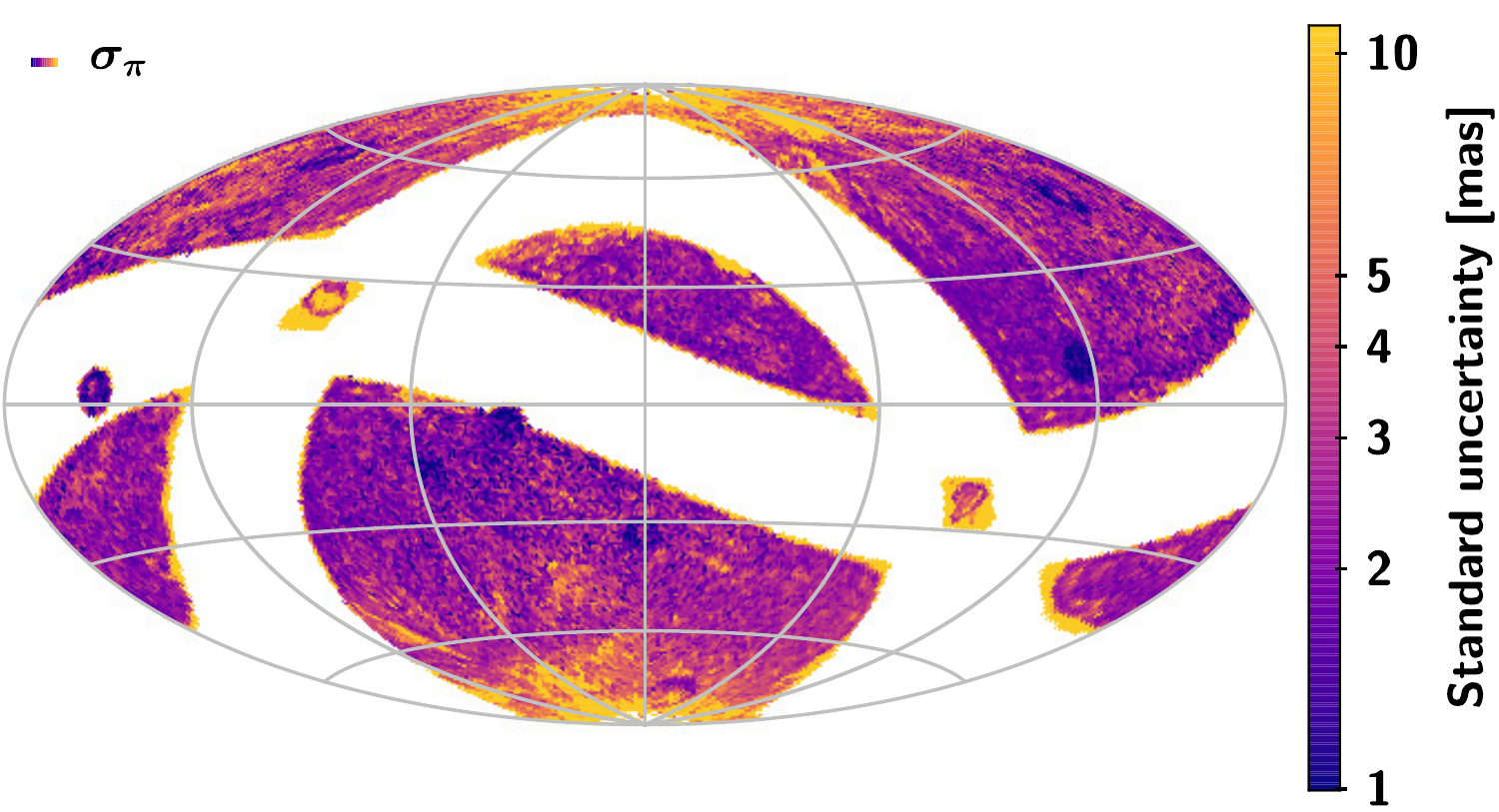}   
         \end{minipage}
         \\
         \\
         \\
         \begin{minipage}{0.2\linewidth}
               \centering
               \includegraphics[scale=0.3]{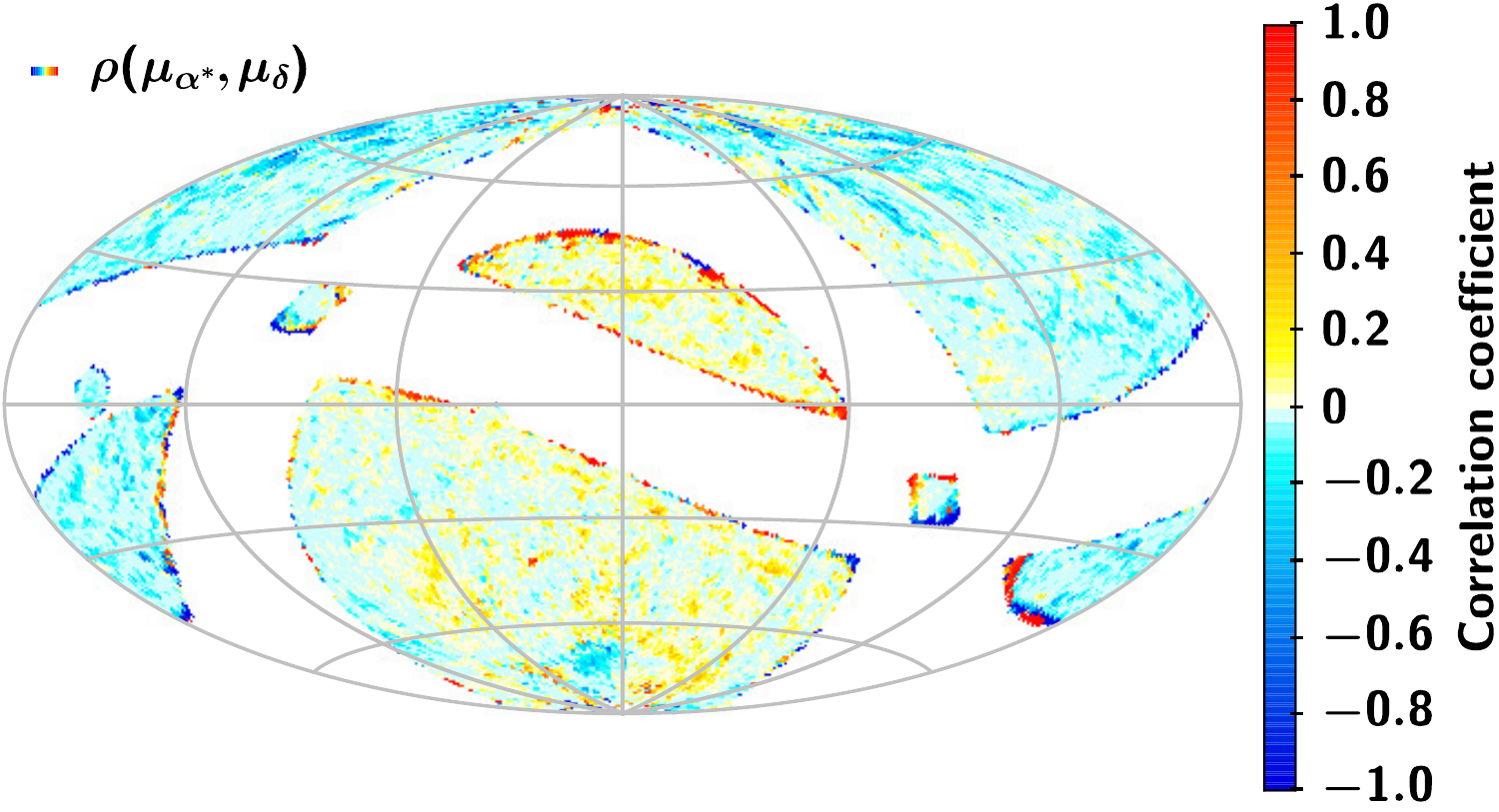}   
         \end{minipage}%
         \begin{minipage}{0.2\linewidth}
               \centering
               \includegraphics[scale=0.3]{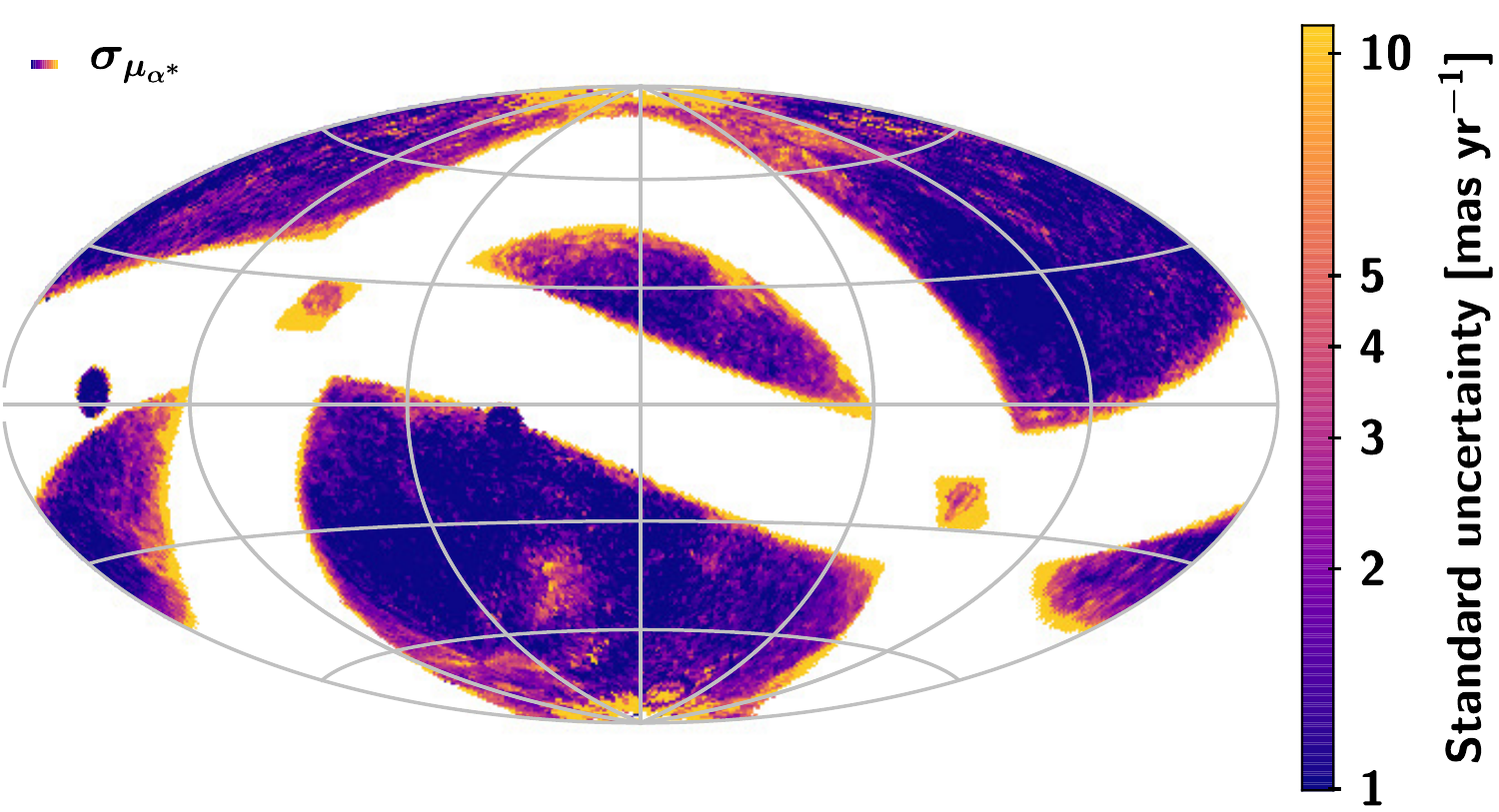}   
         \end{minipage}
         \\
         \\
         \\
         \begin{minipage}{0.2\linewidth}
               \centering
               \includegraphics[scale=0.3]{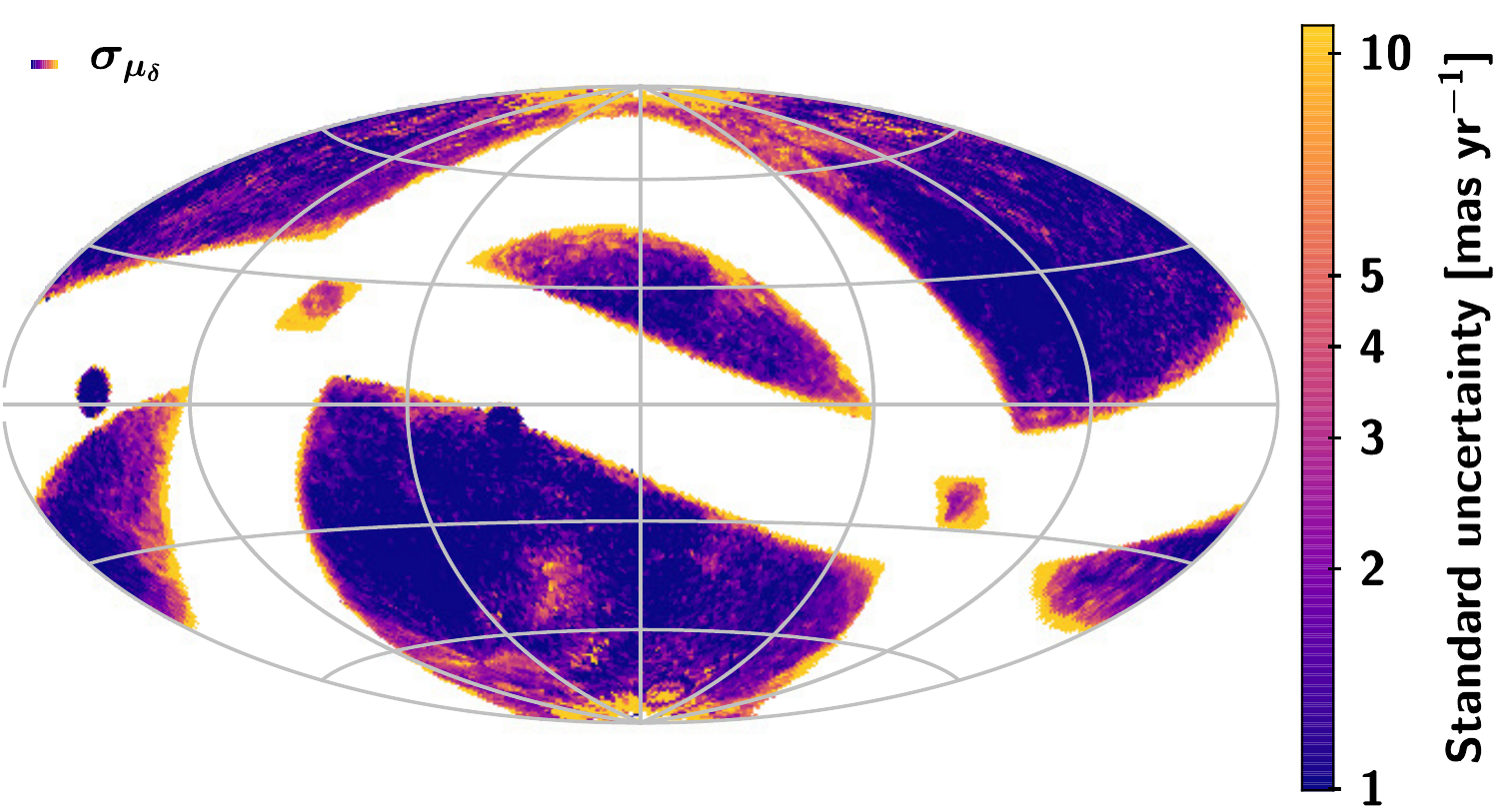}   
         \end{minipage}
     \caption{Summary statistics for all sources in the final catalogue. The five maps along the main diagonal show, from top-right to bottom-left, the standard uncertainties in $\alpha^{\ast}$, $\delta$, $\pi$, $\mu_{\alpha^{\ast}}$, $\mu_\delta$. The ten maps above the diagonal show the correlation coefficients, in the range $-1$ to $+1$, between the corresponding parameters on the main diagonal. All maps use an Aitoff projection in equatorial coordinates, with origin $\alpha=\delta=0$ at the centre and $\alpha$ increasing from right to left. Median values are shown in cells of about 0.84 $deg^2$.}
      \label{bzbqddqtq}
      \end{figure}
   \end{landscape}

    Table~\ref{ZTJSJD} and Fig.~\ref{Vbaifenshu} summarize the results of standard uncertainties (subdivided by magnitude) for the astrometric parameters in the final catalogue. According to the overall calculation of the astrometric parameters, the accuracy of parallax and proper motion of CSST is about 0.1 to 1.0 mas ($\cdot$ yr$^{-1}$) for the sources of 18-22 mag, and 1 to 10 mas ($\cdot$ yr$^{-1}$) for the sources of 22-26 mag, respectively. Fig.~\ref{bzbqddqtq} shows the distribution of standard uncertainties and correlation coefficients of the astrometric parameters in the final catalogue. The correlations between the astrometric parameters are pretty strong (up to $\pm$1) in the areas where the observation times are too concentrated. Similar situations can be found for the uncertainties of the parameters. This is predictable and reflects the inadequacy of current sky survey strategies.
    %
	\begin{table}[htbp]
	\tiny
 	 \centering
  	\caption{Summary statistics for all sources in the final catalogue.}
   	 \begin{tabular}{lrrrrrrrrl}
   	 \toprule
   	 \toprule
          	& \multicolumn{8}{c}{Value at g =}                           &  \\
   	 Quantity & 18-19 & 20    & 21    & 22    & 23    & 24    & 25    & 26    & Unit \\
   	 \midrule
    Median standard uncertainty in $\Delta\alpha^{\ast}$ & 0.416  & 0.621  & 0.938  & 1.410  & 2.157  & 3.361  & 5.759  & 12.631 & mas \\
        Median standard uncertainty in $\Delta\delta$ & 0.343  & 0.518  & 0.780  & 1.175  & 1.792  & 2.793  & 4.784  & 10.404 & mas \\
        Median standard uncertainty in $\pi$ & 0.196  & 0.292  & 0.441  & 0.663  & 1.015  & 1.589  & 2.727  & 5.990 & mas \\
        Median standard uncertainty in $\mu_{\alpha^{\ast}}$ & 0.181  & 0.273  & 0.407  & 0.615  & 0.938  & 1.459  & 2.486  & 5.268 & mas/yr \\
        Median standard uncertainty in $\mu_\delta$ & 0.160  & 0.245  & 0.361  & 0.549  & 0.837  & 1.301  & 2.219  & 4.711 & mas/yr \\
      \bottomrule
   	 \end{tabular}%
  	\label{ZTJSJD}%
	\end{table}%
\section{Discussion}\label{5jie}
   \subsection{Suggestions for optimizing survey schedule}
   In this section, we evaluate the effectiveness of the survey strategy used during the simulation by analyzing the astrometric capability of CSST and give preliminary suggestions for optimizing the survey strategy from an astrometric perspective.
   \par
\begin{figure}[htbp]
      \begin{minipage}{0.33\linewidth}
         \centering
         \includegraphics[scale=0.37]{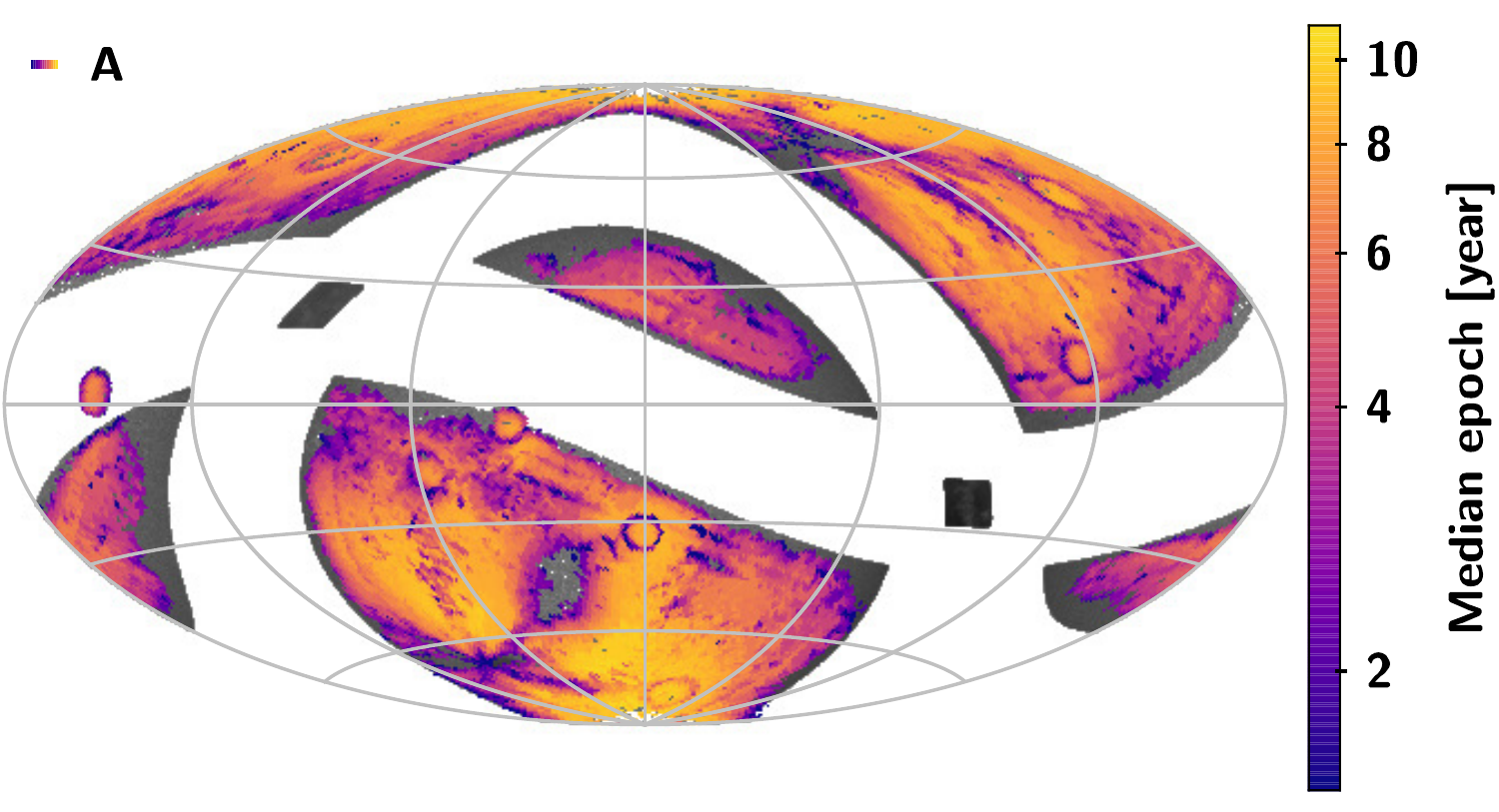}
      \end{minipage}%
      \begin{minipage}{0.33\linewidth}
         \centering
         \includegraphics[scale=0.37]{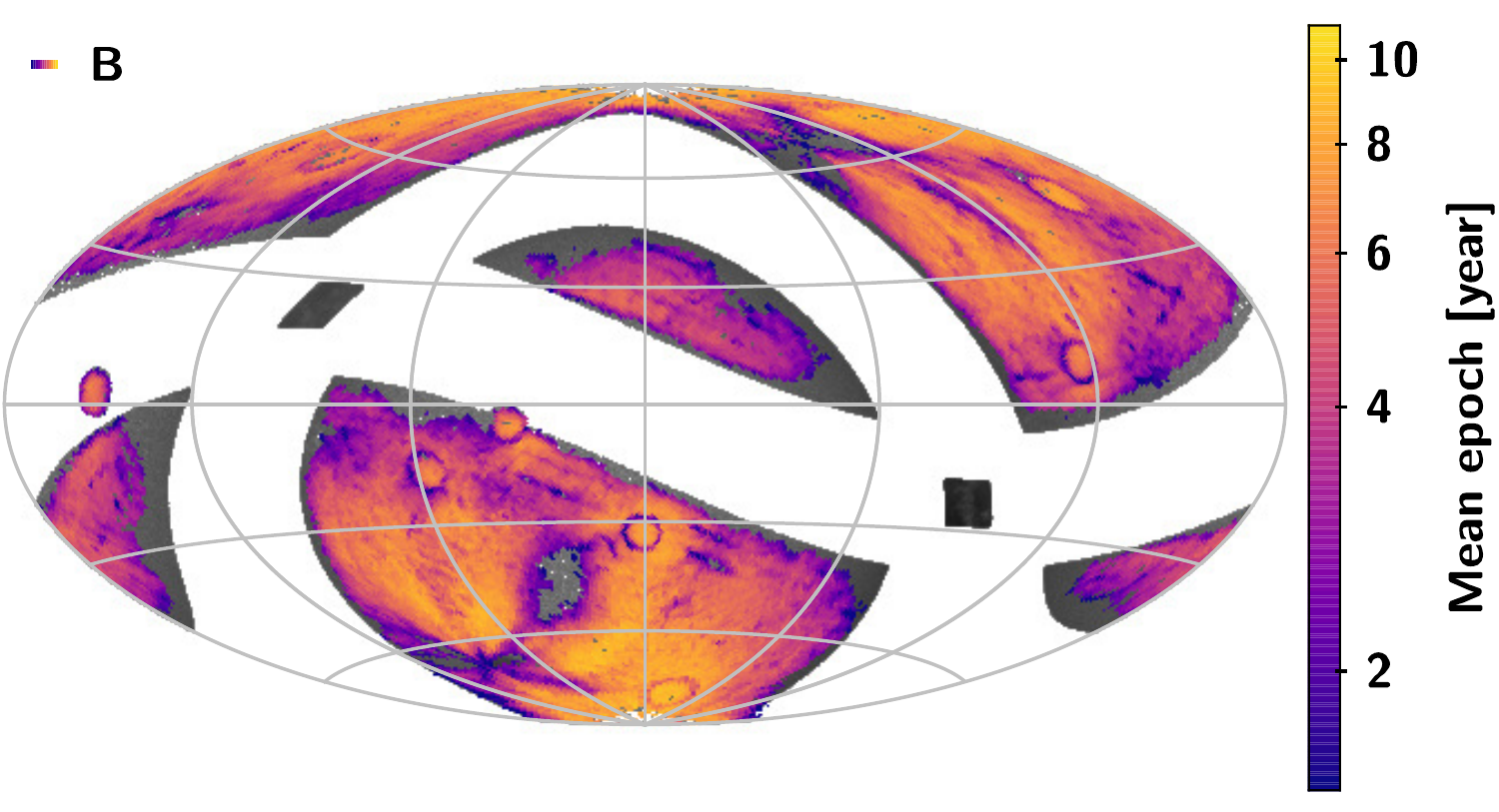}
      \end{minipage}%
      \begin{minipage}{0.33\linewidth}
         \centering
         \includegraphics[scale=0.37]{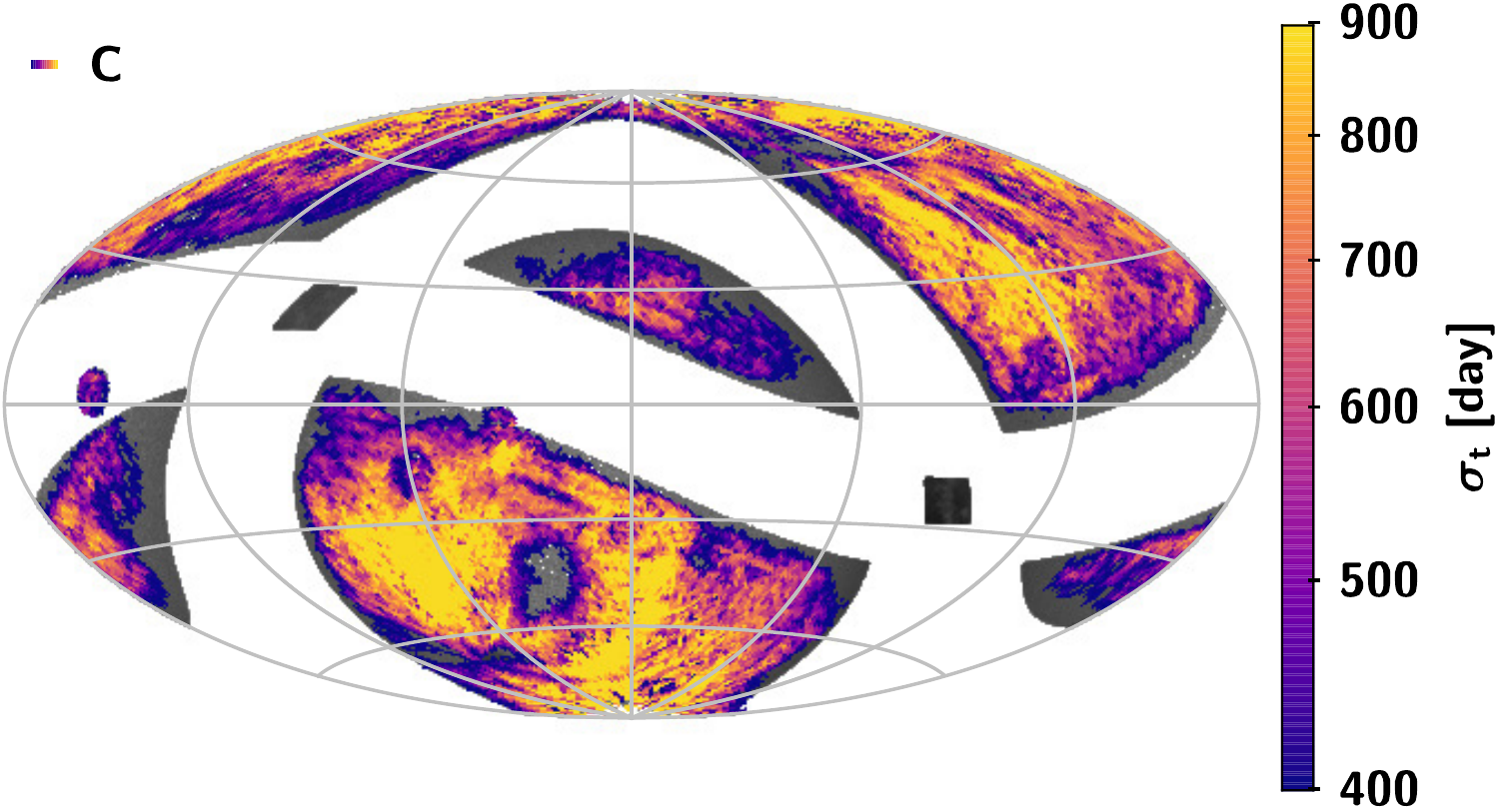}
      \end{minipage}
      
      \begin{minipage}{0.33\linewidth}
         \centering
         \includegraphics[scale=0.37]{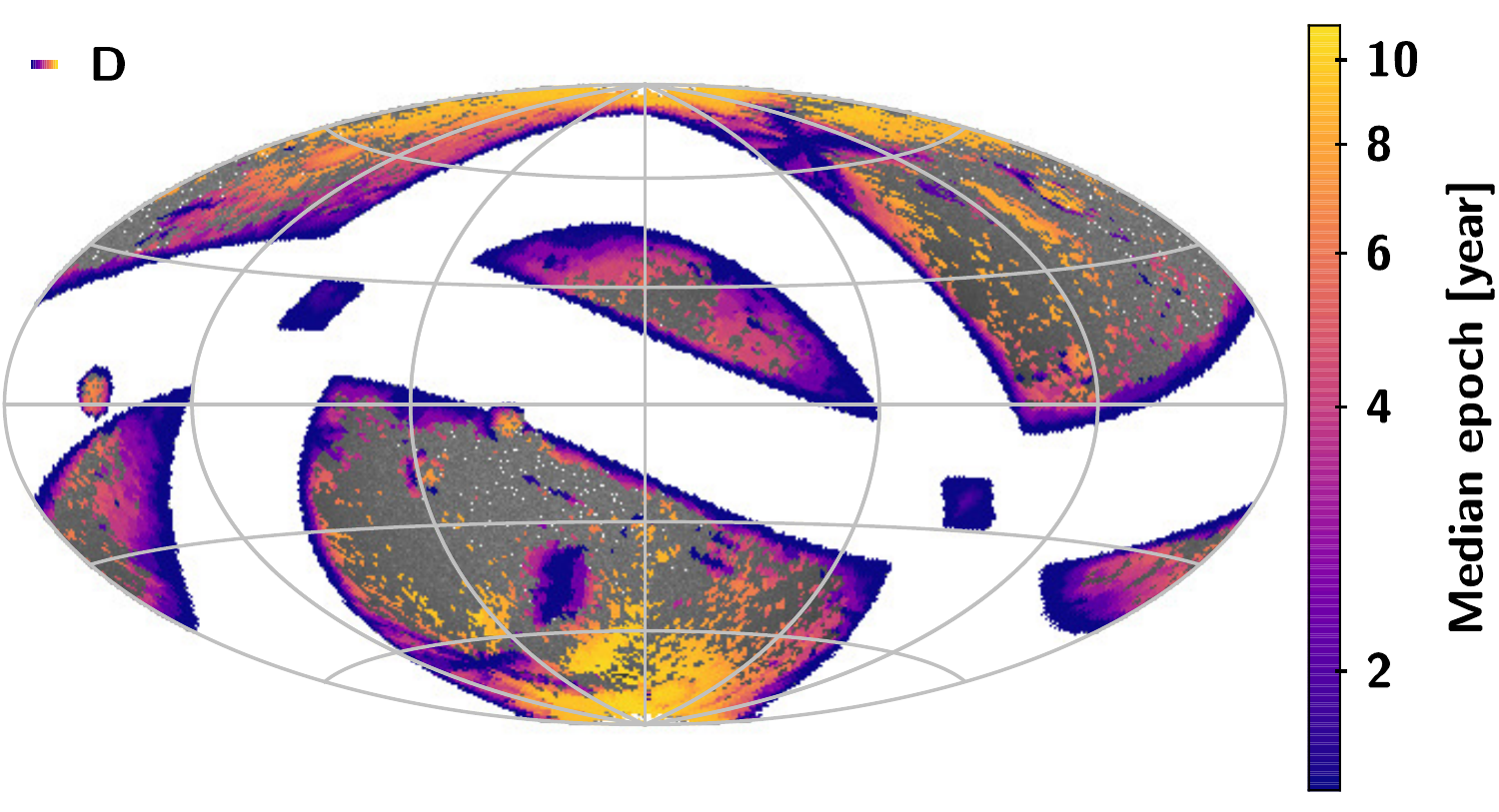}
      \end{minipage}%
      \begin{minipage}{0.33\linewidth}
         \centering
         \includegraphics[scale=0.37]{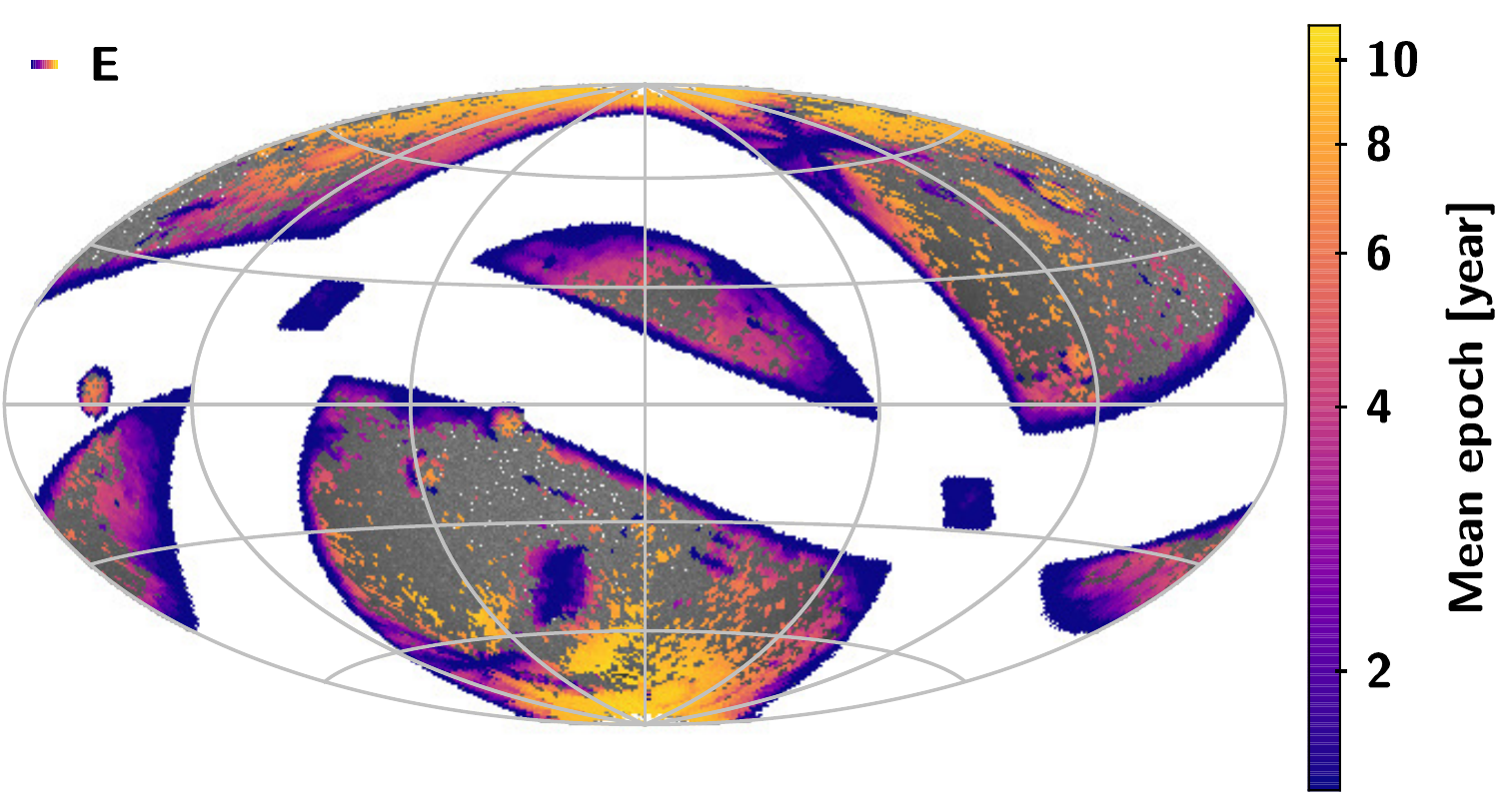}
      \end{minipage}%
      \begin{minipage}{0.33\linewidth}
         \centering
         \includegraphics[scale=0.37]{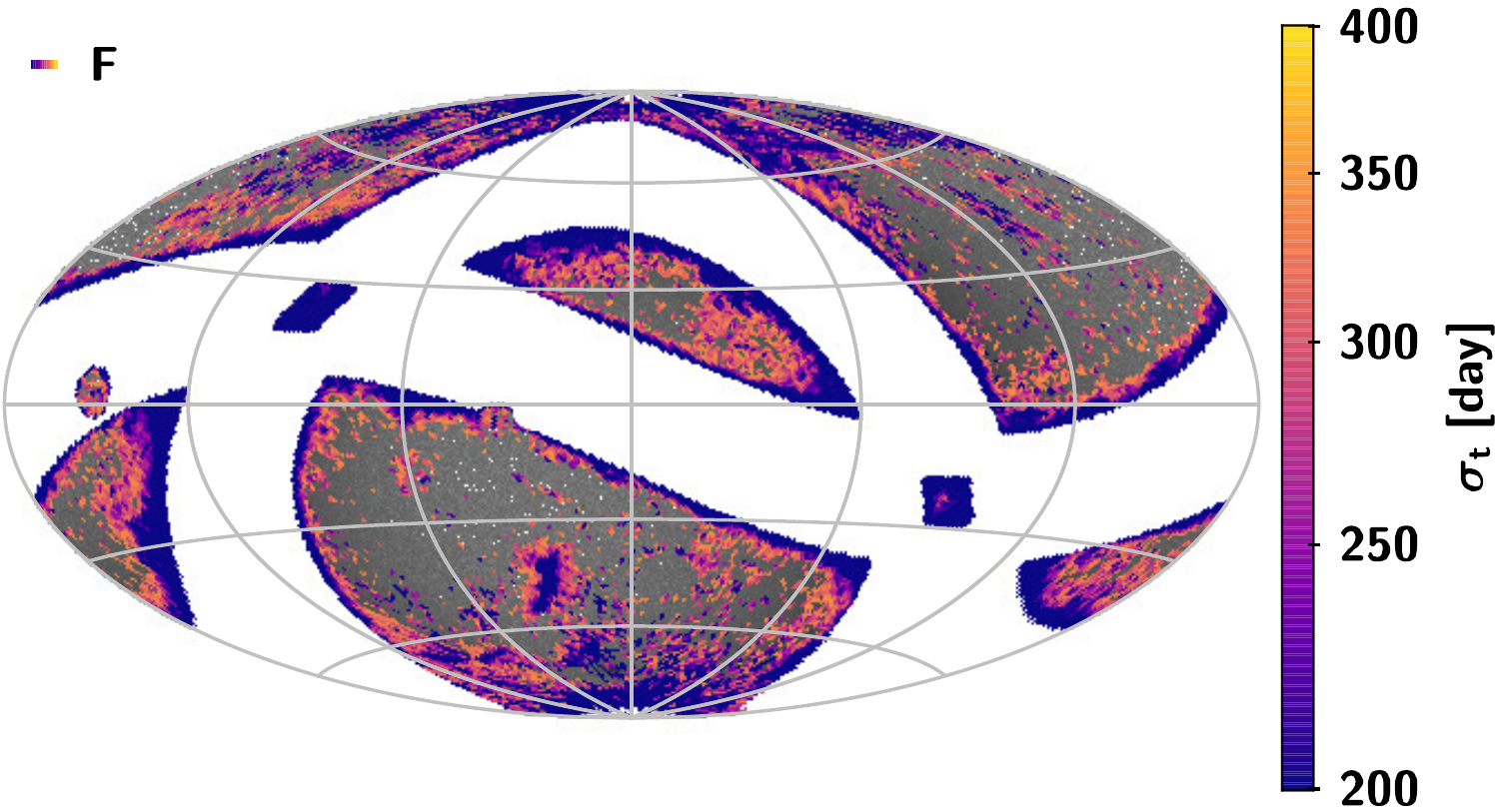}
      \end{minipage}%
      
      \caption{Distributions of the observation time intervals: $\mathbf{A})$ median epoch of the sources with $\sigma_t>\sigma_{mean}$; $\mathbf{B})$ mean epoch of the sources with $\sigma_t>\sigma_{mean}$; $\mathbf{C})$ $\sigma_t$ of the sources with $\sigma_t>\sigma_{mean}$; $\mathbf{D})$ median epoch of the sources with $\sigma_t<\sigma_{mean}$; $\mathbf{E})$ mean epoch of the sources with $\sigma_t<\sigma_{mean}$; $\mathbf{F})$ $\sigma_t$ of the sources with $\sigma_t<\sigma_{mean}$. The gray dots represent all the sources in the final catalogue and are used as reference background stars. All maps use an Aitoff projection in equatorial coordinates, with origin $\alpha=\delta=0$ at the centre and $\alpha$ increasing from right to left. Median values are shown in cells of about 0.84 $deg^2$.}
      \label{csstxtbp2}%
    \end{figure}
\begin{figure*}[htbp]
         \centering
         \includegraphics[scale=0.7]{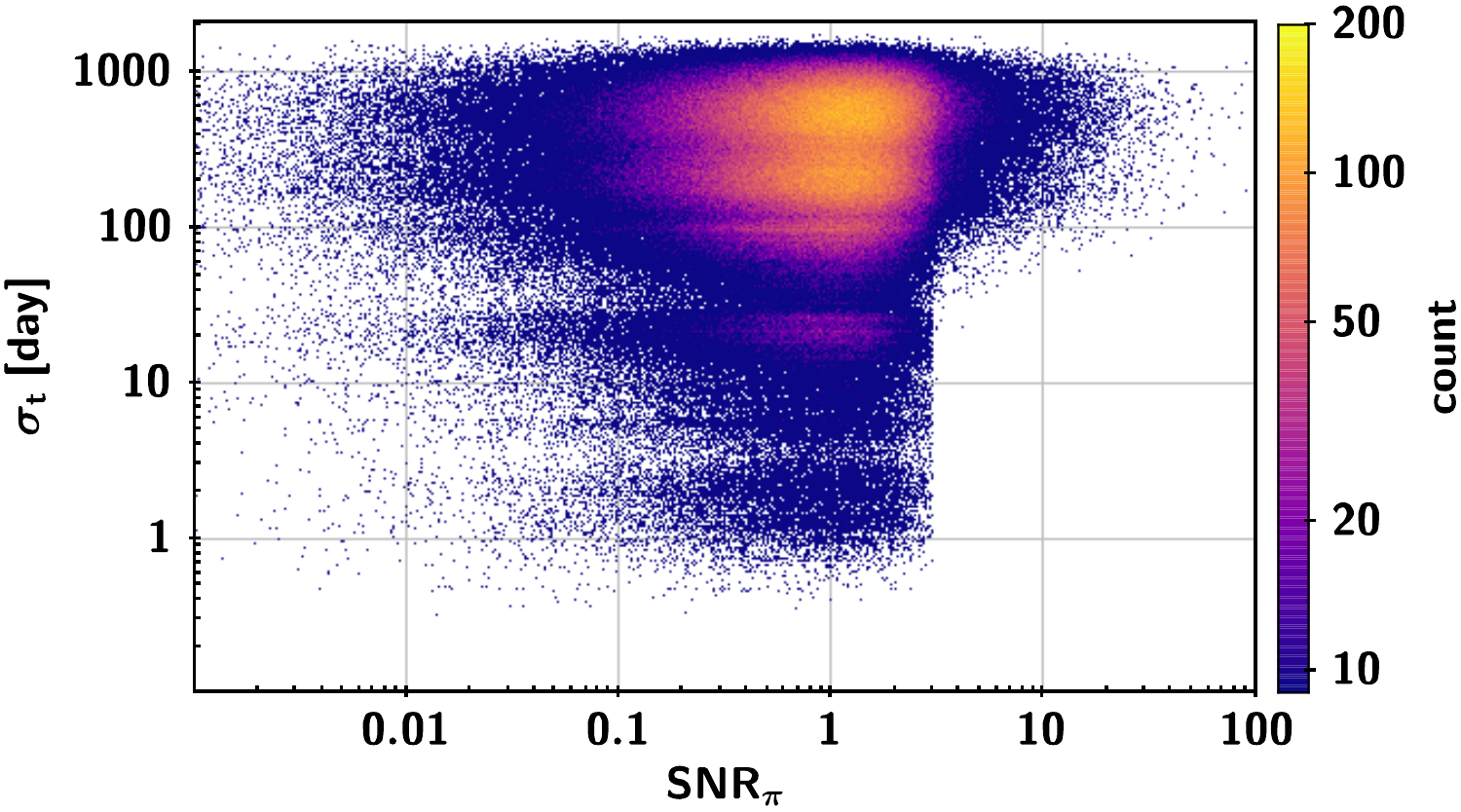}
      \caption{Relationship between the dispersion of observation series and signal-to-noise ratio of parallax for the parallax solvable sources in the final catalogue.}
      \label{snr_up_T}%
    \end{figure*}

The number and the distribution of the astrometric epoch data are the keys to obtaining high-quality astrometric parameters of the sources. The statistical information of the sources with $\sigma_t>\sigma_{mean}$ ($\sigma_{mean}$=349 days) and $\sigma_t<\sigma_{mean}$ are shown in Table~\ref{sigma_t_good} and Table~\ref{sigma_t_bad}, respectively. Fig.~(\ref{csstxtbp2}) shows the distributions of $\sigma_t$, median and mean epoch of the observation time intervals. A small $\sigma_t$ suggests that the observation time is too concentrated, which will lead to an unfavorable solution of the astronomical parameters, as shown in Fig.~(\ref{snr_up_T}). The distribution of the sources with $\sigma_t<\sigma_{mean}$ is mainly near the edge of the galactic equator and the intersection of the galactic equator and the ecliptic, and some of them are located in the high declination region and the edge of the ecliptic. These regions are corresponding to the regions with large uncertainty in proper motion in Fig.~\ref{bzbqddqtq}. By comparing the results of Table~\ref{ZTJSJD}, Table~\ref{sigma_t_good}, and Table~\ref{sigma_t_bad}, it can be found that optimizing the survey schedule in these regions may improve the quality and quantity of parameter solutions of the corresponding celestial bodies, thereby enhancing the astrometric capability of CSST. We suggest that these regions be scheduled for more even observations in the future during the optimization of the survey strategy.
\begin{table}[htbp]
  \centering
    \tiny
  \caption{Summary statistics for the sources with $\sigma_t > \sigma_{mean}$ in the final catalogue.}
    \begin{tabular}{lrrrrrrrrl}
    \toprule
    \toprule
          & \multicolumn{8}{c}{Value at g =}                           &  \\
    Quantity & 18-19 & 20    & 21    & 22    & 23    & 24    & 25    & 26    & Unit \\
    \midrule
    Median standard uncertainty in $\Delta\alpha^{\ast}$ & 0.341  & 0.515  & 0.771  & 1.167  & 1.775  & 2.762  & 4.736  & 10.229 & mas \\
        Median standard uncertainty in $\Delta\delta$ & 0.323  & 0.486  & 0.734  & 1.107  & 1.685  & 2.620  & 4.497  & 9.710 & mas \\
        Median standard uncertainty in $\pi$ & 0.163  & 0.241  & 0.364  & 0.548  & 0.837  & 1.304  & 2.241  & 4.870 & mas \\
        Median standard uncertainty in $\mu_{\alpha^{\ast}}$ & 0.078  & 0.115  & 0.173  & 0.263  & 0.401  & 0.623  & 1.063  & 2.265 & mas/yr \\
        Median standard uncertainty in $\mu_\delta$ & 0.075  & 0.110  & 0.165  & 0.250  & 0.382  & 0.594  & 1.012  & 2.155 & mas/yr \\
      Proportion of the objects with $SNR_\pi>1$ & 76.7\% & 69.7\% & 60.5\% & 50.8\% & 42.3\% & 34.0\% & 27.2\% & 20.8\% & \\
    Proportion of the objects with $SNR_\pi>3$ & 47.8\% & 33.8\% & 23.7\% & 14.7\% & 8.1\% & 3.9\% & 1.4\% & 0.3\% & \\
    \bottomrule
 \end{tabular}%
  \label{sigma_t_good}%
\end{table}%
    \begin{table}[htbp]
  \centering
    \tiny
  \caption{Summary statistics for the sources with $\sigma_t < \sigma_{mean}$ in the final catalogue.}
    \begin{tabular}{lrrrrrrrrl}
    \toprule
    \toprule
          & \multicolumn{8}{c}{Value at g =}                           &  \\
    Quantity & 18-19 & 20    & 21    & 22    & 23    & 24    & 25    & 26    & Unit \\
    \midrule
    Median standard uncertainty in $\Delta\alpha^{\ast}$ & 0.527  & 0.806  & 1.218  & 1.824  & 2.794  & 4.385  & 7.462  & 16.394 & mas \\
        Median standard uncertainty in $\Delta\delta$ & 0.365  & 0.557  & 0.837  & 1.254  & 1.916  & 2.996  & 5.117  & 11.169 & mas \\
        Median standard uncertainty in $\pi$ & 0.259  & 0.386  & 0.587  & 0.871  & 1.336  & 2.117  & 3.618  & 7.955 & mas \\
        Median standard uncertainty in $\mu_{\alpha^{\ast}}$ & 0.356  & 0.532  & 0.822  & 1.209  & 1.852  & 2.942  & 5.046  & 11.092 & mas/yr \\
        Median standard uncertainty in $\mu_\delta$ & 0.284  & 0.442  & 0.668  & 0.996  & 1.508  & 2.384  & 4.069  & 8.880 & mas/yr \\
      Proportion of the objects with $SNR_\pi>1$ & 56.0\% & 48.7\% & 41.7\% & 35.3\% & 29.2\% & 25.1\% & 21.5\% & 18.4\% & \\
    Proportion of the objects with $SNR_\pi>3$ & 24.1\% & 15.5\% & 9.6\% & 5.5\% & 2.7\% & 1.4\% & 0.5\% & 0.1\% & \\
    \bottomrule
\end{tabular}%
  \label{sigma_t_bad}%
\end{table}
   \par
\begin{figure}[htbp]
         \begin{minipage}{0.5\linewidth}
         \centering
         \includegraphics[scale=0.56]{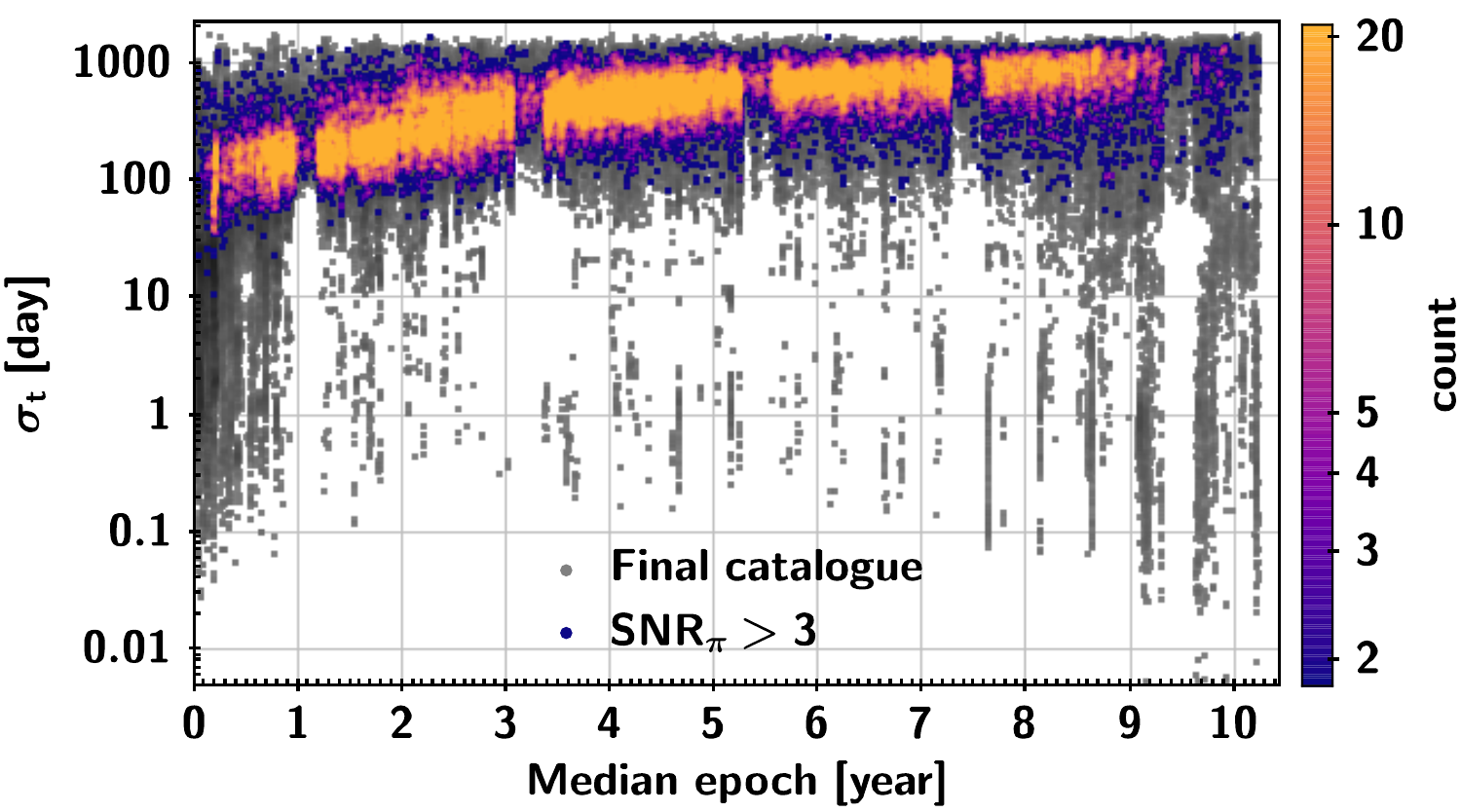}
      \end{minipage}%
      \begin{minipage}{0.5\linewidth}
         \centering
         \includegraphics[scale=0.56]{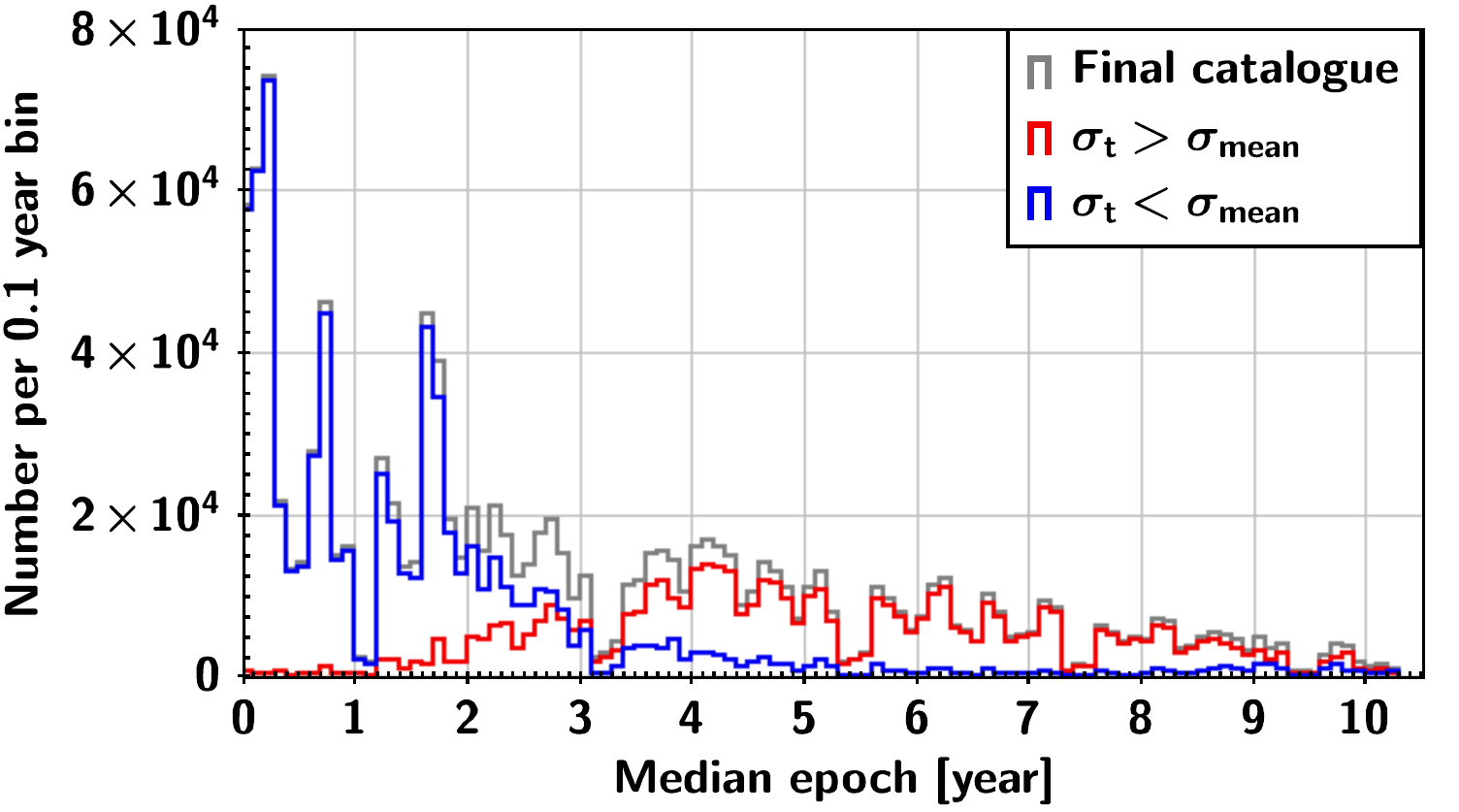}
      \end{minipage}
      
      \begin{minipage}{0.5\linewidth}
         \centering
         \includegraphics[scale=0.56]{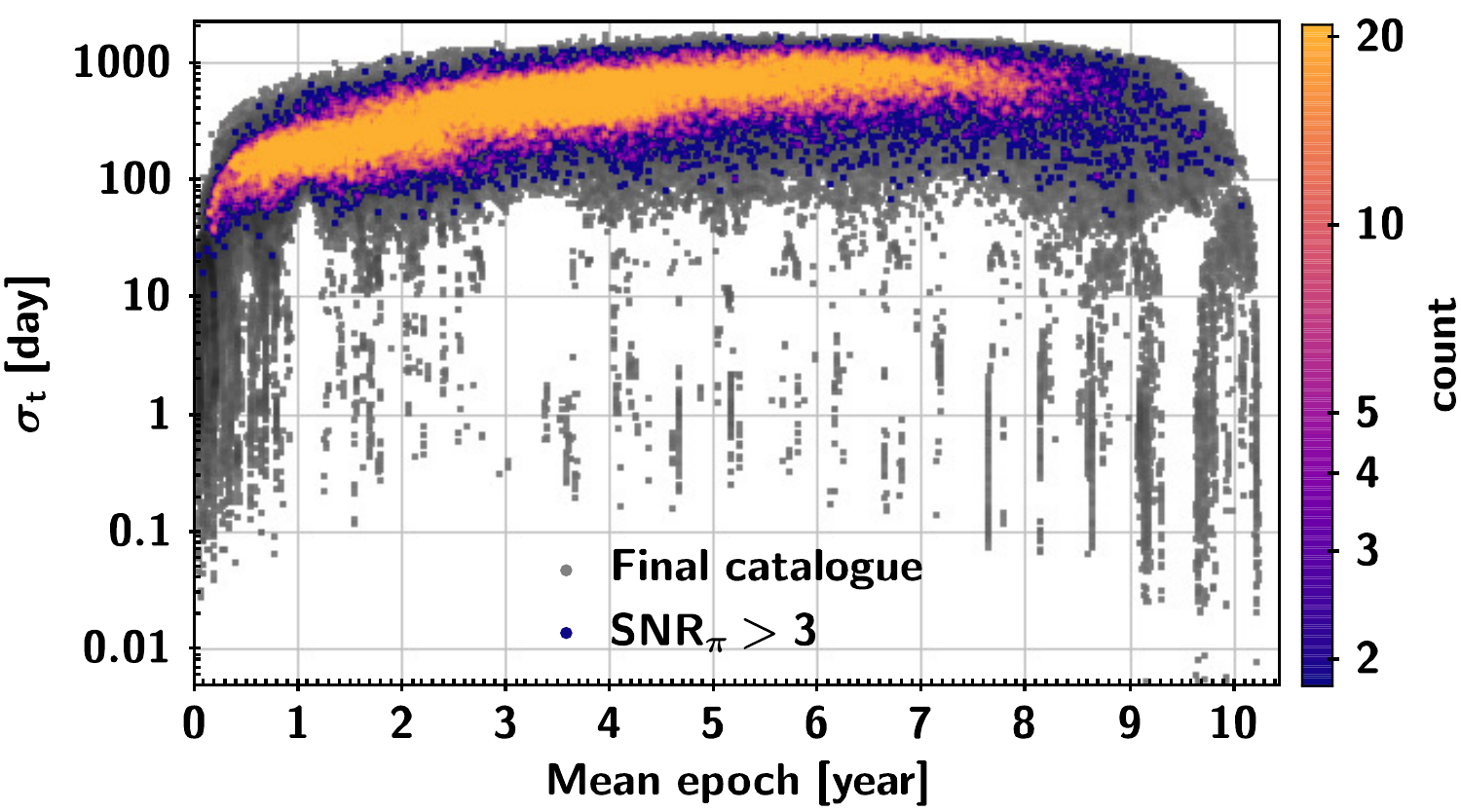}
      \end{minipage}%
      \begin{minipage}{0.5\linewidth}
         \centering
         \includegraphics[scale=0.56]{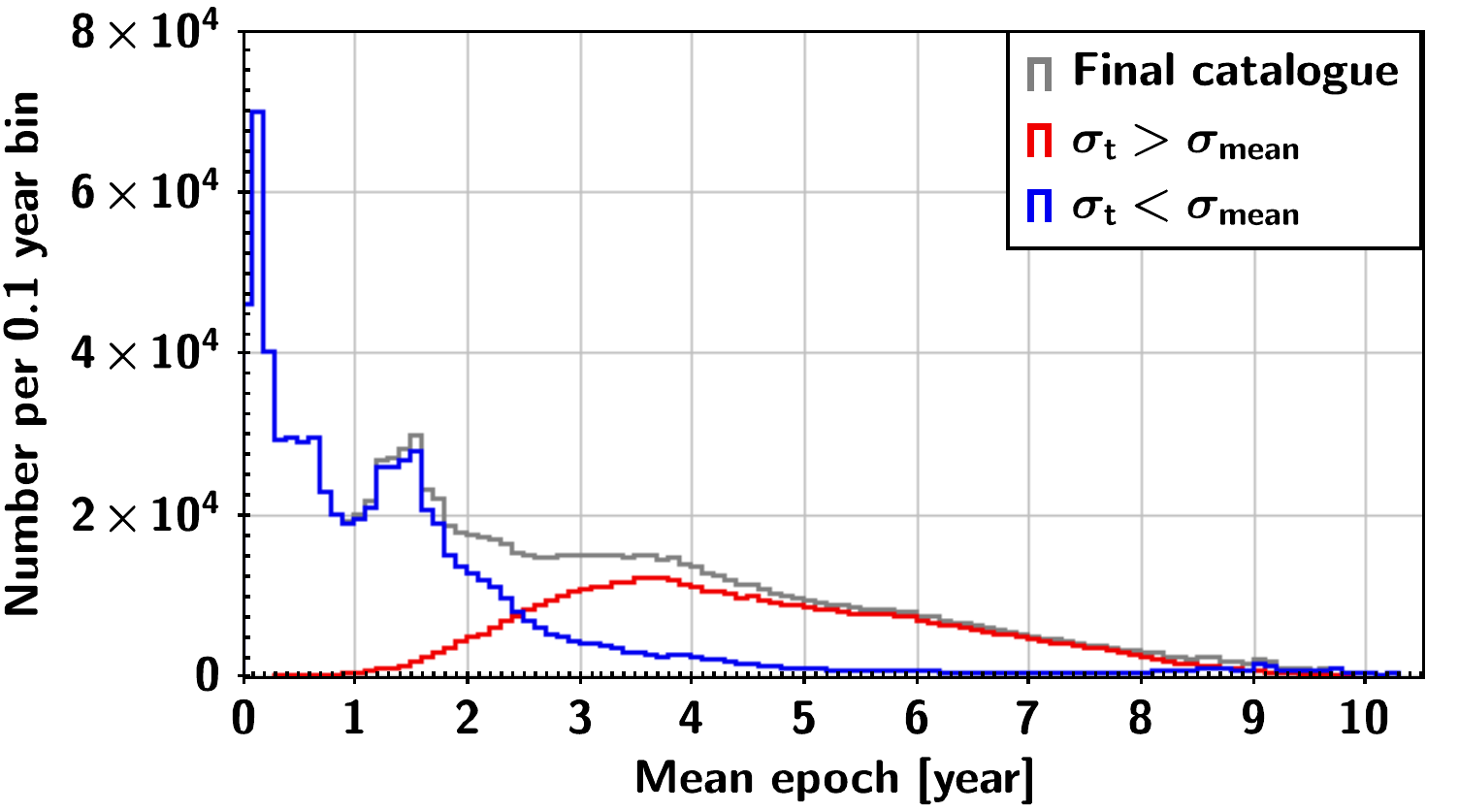}
      \end{minipage}
      
      \begin{minipage}{0.5\linewidth}
         \centering
         \includegraphics[scale=0.56]{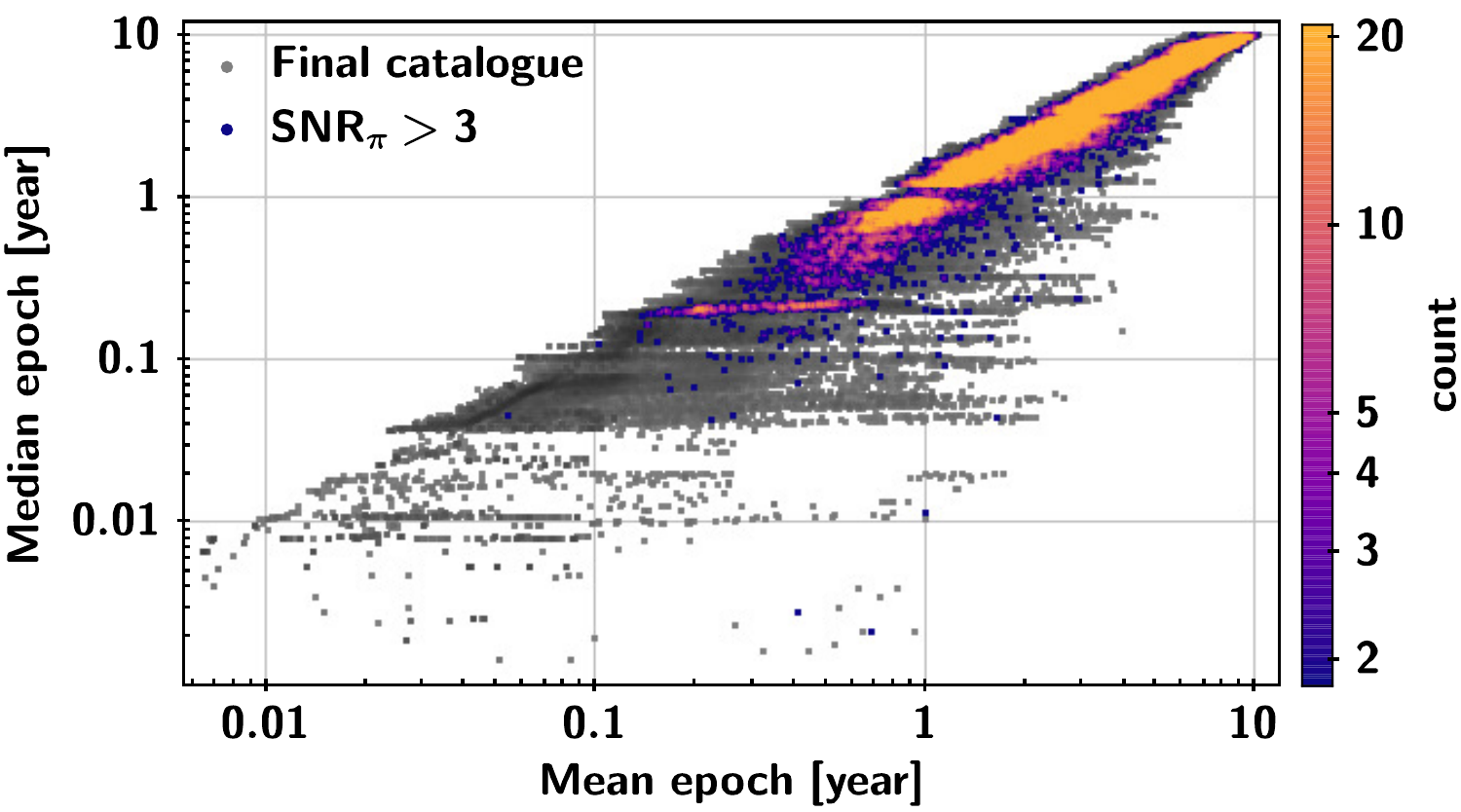}
      \end{minipage}%
      \begin{minipage}{0.5\linewidth}
         \centering
         \includegraphics[scale=0.56]{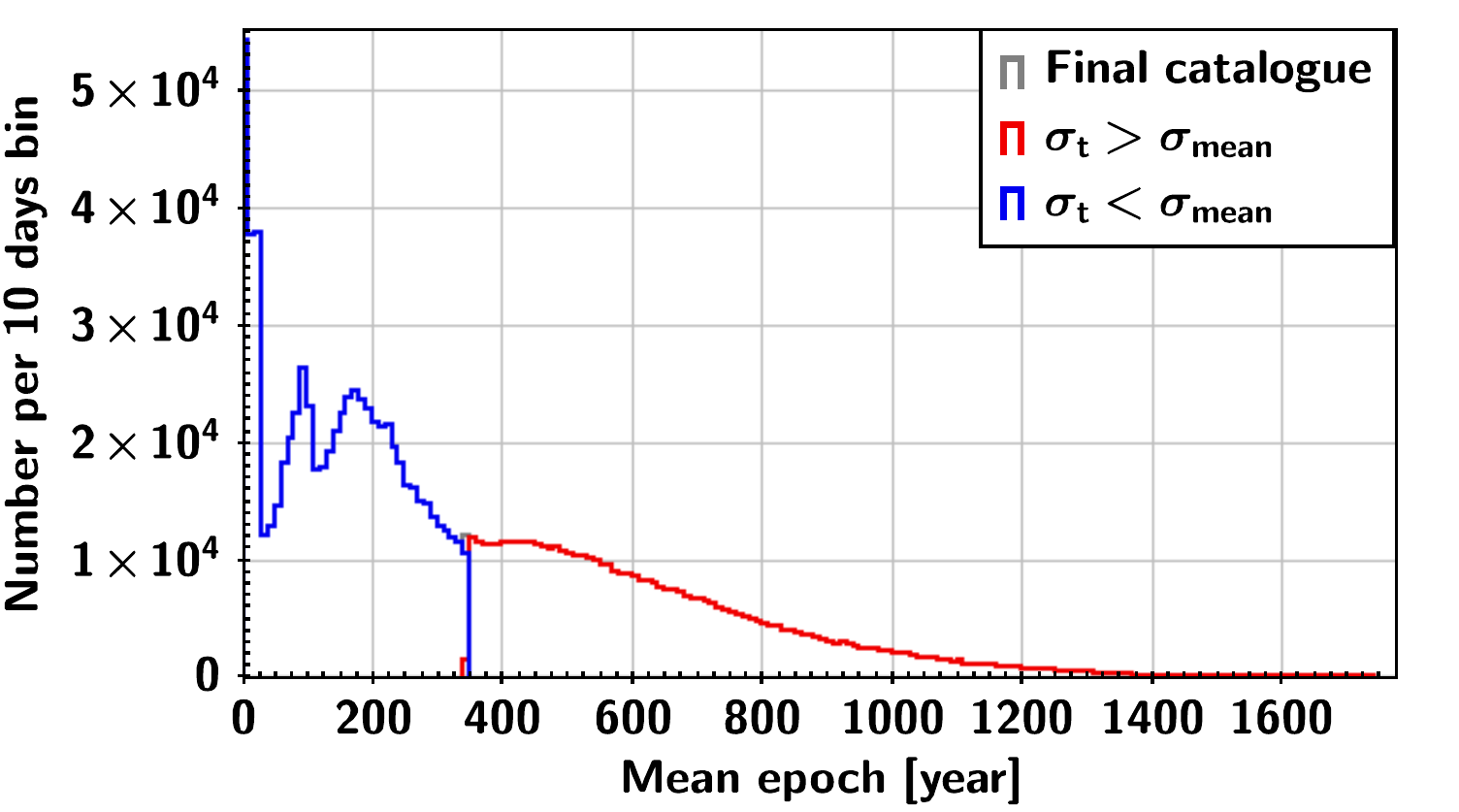}
      \end{minipage}%
      
      \caption{Summary statistics of the celestial observation time series. $Left\ figures:$ The color dots represent the sources with $SNR_\pi>$3. The grey dots represent all the sources in the final catalogue and are used as reference background stars. $Right\ figures:$ The grey steps are all the sources in the final catalogue. The red and blue steps are the sources with $\sigma_t>\sigma_{mean}$ and $\sigma_t<\sigma_{mean}$, respectively.}
      \label{Medians_t}%
    \end{figure}

 Fig.~\ref{Medians_t} shows more directly the observational effects of the survey schedule used in the simulation, where there are a large number of sources with $\sigma_t \ll \sigma_{mean}$ in the first year, and optimizing their observational schedule can add more sources with high-precision astrometric parameters to the CSST observational sample. Combining the simulation of CSST astrometric capability with the optimization of the survey schedule can lead to a more efficient survey schedule and provide a more high-precision data sample for CSST science.
\section{Conclusions}\label{6jie}
We develop an astrometric model to analyze the simulation astrometric epoch data and evaluate the astrometric capability of CSST under a specific survey schedule. The simulated observations are implemented based on a specific survey schedule, in which we simulate the observation errors as Gaussian noise based on the magnitude dependence of astrometric uncertainty.  

In the final catalogue, 97.18\% of the sources have 5-parameter solutions, and 98.57\% of the sources have 4-parameter solutions. The accuracy of parallax and proper motion of CSST is about the order of 0.1 to 1.0 mas ($\cdot$ yr$^{-1}$) for the sources of 18-22 mag in g band, and 1 to 10 mas ($\cdot$ yr$^{-1}$) for the sources of 22-26 mag in g band, respectively. The accuracy of the proper motion of CSST is significantly better for the sources with $\sigma_t>\sigma_{mean}$ compared to the sources with $\sigma_t<\sigma_{mean}$. The assumptions used in this paper are very optimistic and simple, it is foreseen that the results from the real survey could be worse. We will study these issues in detail in the future.
   \par
Through the analysis of the astrometric capability of CSST, we pointed out aspects of the specific survey schedule that can be optimized. The quality and quantity of CSST observations can be further improved by optimizing the observation arrangements in the sky areas where the observations do not meet the requirements. In the future, we will iterate with the survey schedule mutually to provide independent evaluation opinions for CSST's survey strategic arrangement.

\section*{Author Contributions}
S-LL and Z-XQ are responsible for supervising the model construction and astrometric epoch data simulation. Z-SF simulates the CSST data and solves the observations equation with meticulous efforts and wrote the manuscript with help mainly from S-LL. Besides, X-YP, Y-Y, and Q-QW contributed to the physical interpretation and discussion. SL and Y-HX contributed to the formation of the CSST survey schedule.

\section*{Funding}
This work has been supported by the Youth Innovation Promotion Association CAS with Certificate Number 2022259, the grants from the Natural Science Foundation of Shanghai through grant 21ZR1474100, and National Natural Science Foundation of China (NSFC) through grants 12173069, and 11703065. We acknowledge the science research grants from the China Manned Space Project with NO.CMS-CSST-2021-A12, NO.CMS-CSST-2021-B10 and NO.CMS-CSST-2021-B04.

\section*{Acknowledgments}
This work has made use of data from the European Space Agency (ESA) mission Gaia (https://www.cosmos.esa.int/gaia), processed by the Gaia Data Processing and Analysis Consortium (DPAC, https://www.cosmos.esa.int/web/gaia/ dpac/consortium). Funding for the DPAC has been provided by national institutions, in particular the institutions participating in the Gaia Multilateral Agreement. We are also very grateful to the developers of the TOPCAT \citep{taylor2005topcat} software.

\bibliographystyle{Frontiers-Harvard} 

\bibliography{test}

\begin{appendices}
  \setcounter{equation}{0}
  \renewcommand{\theequation}{A.\arabic{equation}}
    \section{Solving the observation error equation for a single star}\label{1app}
    In this section, the procedure of constructing and solving the observation error equation containing the matrix of astrometric parameters is presented. In Sect.~\ref{2.2jie}, the single observation equation (Eqs.~(\ref{LPCgs1},\ref{LPCgs2})) for a single star is introduced, which has the matrix form: 
    \begin{equation}
      L(t_\mathrm{i}) + v_\mathrm{i} = B(t_\mathrm{i})X \,,
    \end{equation}
where $L(t_\mathrm{i})$ is the observation matrix \citep[Sect. 19.2]{van2013astrometry}: 
    \begin{equation}
      \label{GCFCLBJ3}
      L(t_\mathrm{i}) = 
        \begin{bmatrix}
         \xi(t_\mathrm{i}) \\
         \eta(t_\mathrm{i}) 
        \end{bmatrix} = 
        \begin{bmatrix}
         \frac{\cos \delta(t_\mathrm{i}) \sin(\alpha(t_\mathrm{i})-\alpha_\mathrm{ep})}{\sin \delta_\mathrm{ep} \sin \delta(t_\mathrm{i}) + \cos \delta_\mathrm{ep} \cos \delta(t_\mathrm{i}) \cos(\alpha(t_\mathrm{i})-\alpha_\mathrm{ep})} \\
         \frac{\cos \delta_\mathrm{ep} \sin \delta(t_\mathrm{i}) - \sin \delta_\mathrm{ep} \cos \delta(t_\mathrm{i}) \cos(\alpha(t_\mathrm{i})-\alpha_\mathrm{ep})}{\sin \delta_\mathrm{ep} \sin \delta(t_\mathrm{i}) + \cos \delta_\mathrm{ep} \cos \delta(t_\mathrm{i}) \cos(\alpha(t_\mathrm{i})-\alpha_\mathrm{ep})}
        \end{bmatrix} \,; 
   \end{equation}
$v_\mathrm{i}$ is the correction value of $L(t_\mathrm{i})$, indicating that there are slight observation errors in $\xi(t_\mathrm{i})$ and $\eta(t_\mathrm{i})$; $X$ is the unknown parameter matrix, which is defined as:
   \begin{equation}
      \label{GCFCLBJ3}
      X = 
        \begin{bmatrix}
         \Delta \alpha^{\ast}&
         \Delta \delta&
         \mu_{\alpha^{\ast}}&
         \mu_\delta&
         \pi
        \end{bmatrix}^\mathrm{T} \,;
   \end{equation}
$B(t_\mathrm{i})$ is the coefficient matrix, which is defined as:
   \begin{equation}
      \label{GCFCLBJ2}
         B(t_\mathrm{i}) =
         \begin{bmatrix} 
            1 & 0 & \frac{t_\mathrm{i}-t_\mathrm{ep}}{1-\mathbf{r_\mathrm{ep}^\prime}\mathbf{b_\mathrm{O}}(t_\mathrm{i})\pi /Au} & 0 & \frac{-\mathbf{p_\mathrm{ep}^\prime} \mathbf{b_\mathrm{O}}(t_\mathrm{i}) /Au}{1-\mathbf{r_\mathrm{ep}^\prime}\mathbf{b_\mathrm{O}}(t_\mathrm{i})\pi /Au} \\
            0 & 1 & 0 & \frac{t_\mathrm{i}-t_\mathrm{ep}}{1-\mathbf{r_\mathrm{ep}^\prime}\mathbf{b_\mathrm{O}}(t_\mathrm{i})\pi /Au} & \frac{-\mathbf{q_\mathrm{ep}^\prime}\mathbf{b_\mathrm{O}}(t_\mathrm{i}) /Au}{1-\mathbf{r_\mathrm{ep}^\prime}\mathbf{b_\mathrm{O}}(t_\mathrm{i})\pi /Au} 
         \end{bmatrix} \,.
   \end{equation}
   \par
The error equation is established by associating the single-star observation equation for n different moments:
    \begin{equation}
      \label{NCGCFCZ2}
       \begin{bmatrix}v_\mathrm{1}\\v_\mathrm{2}\\\cdots\\v_\mathrm{i}\\\cdots\\v_\mathrm{n} 
       \end{bmatrix}
       =
       \begin{bmatrix} B(t_\mathrm{1})\\B(t_\mathrm{2})\\\cdots\\B(t_\mathrm{i})\\\cdots\\B(t_\mathrm{n}) 
       \end{bmatrix}X -
       \begin{bmatrix}L(t_\mathrm{1})\\L(t_\mathrm{2})\\\cdots\\L(t_\mathrm{i})\\\cdots\\L(t_\mathrm{n}) 
       \end{bmatrix}
    \end{equation}
or 
    \begin{equation}
      \label{NCGCFCZ3}
       V=BX-L \,.
    \end{equation}
When the matrix form of the error equation of the single star is established, the expression of the unknown parameter matrix can be obtained from the least-squares method(the weight matrix of the observation value is the identity matrix):
    \begin{equation}
       \label{ZXECJ}
       X=(B^\mathrm{T}B)^{-1}B^\mathrm{T}L \,.
    \end{equation}
    \par
The iterative equations for solving the unknown parameter matrix can be constructed by the definition Eq.~(\ref{GCFCLBJ3}) and the expression Eq.~(\ref{ZXECJ}). When the iterative equations converge, we can calculate the specific value of the unknown parameter matrix $X$, to obtain the position, parallax, and proper motion of the single star. 
\par
The mean square error ($\sigma_\mathrm{o}$) of the measured value $L$ is:
\begin{equation}
       \label{DWQFC1}
       \sigma_\mathrm{o}=\sqrt{\frac{(BX-L)^\mathrm{T}(BX-L)}{2n-5}}\,.
    \end{equation}
The covariance matrix ($Q_{XX}$) of $X$ is:
\begin{equation}
       \label{DWQFC2}
       Q_{XX} = (B^\mathrm{T}B)^{-1}\,.
    \end{equation}
The variance matrix ($D_{XX}$) of $X$ is:
 \begin{equation}
       \label{XYSZQ3}
       D_{XX} = \sigma_\mathrm{o}^2 Q_{XX}
       =\begin{bmatrix}
          \sigma_{\alpha^{\ast}}^2&\sigma_{\alpha^{\ast} \delta}&\sigma_{\alpha^{\ast} \mu_{\alpha^{\ast}}}&\sigma_{\alpha^{\ast} \mu_\delta}&\sigma_{\alpha^{\ast} \pi}\\
          \sigma_{\delta \alpha^{\ast}}&\sigma_{\delta}^2&\sigma_{\delta \mu_{\alpha^{\ast}}}&\sigma_{\delta \mu_\delta}&\sigma_{\delta \pi}\\
          \sigma_{\mu_{\alpha^{\ast}} \alpha^{\ast}}&\sigma_{\mu_{\alpha^{\ast}} \delta}&\sigma_{\mu_{\alpha^{\ast}}}^2&\sigma_{\mu_{\alpha^{\ast}} \mu_\delta}&\sigma_{\mu_{\alpha^{\ast}} \pi}\\
          \sigma_{\mu_\delta \alpha^{\ast}}&\sigma_{\mu_\delta \delta}&\sigma_{\mu_\delta \mu_{\alpha^{\ast}}}&\sigma_{\mu_\delta}^2&\sigma_{\mu_\delta \pi}\\
          \sigma_{\pi \alpha^{\ast}}&\sigma_{\pi \delta}&\sigma_{\pi \mu_{\alpha^{\ast}}}&\sigma_{\pi \mu_\delta}&\sigma_{\pi}^2
       \end{bmatrix}\,.
    \end{equation}
Thus, the standard uncertainties of the astrometric parameters and the correlation coefficients between each parameter can be obtained:
 \begin{eqnarray}
       \sigma_\mathrm{i} &=& \sqrt{\sigma_\mathrm{i}^2}
       \ \ \ \ \ \ \ \ \ (i=\alpha^{\ast},\ \delta,\ \mu_{\alpha^{\ast}},\ \mu_\delta,\ \pi)
       \\
       \rho_{\mathrm{i} \mathrm{j}} &=& \frac{\sigma_{\mathrm{i} \mathrm{j}}}{\sqrt{\sigma_\mathrm{i}^2 \sigma_\mathrm{j}^2}}
       \ \ \ \ \ (i,j=\alpha^{\ast},\ \delta,\ \mu_{\alpha^{\ast}},\ \mu_\delta,\ \pi\ and\  i\neq j)
       \,.
    \end{eqnarray}
    \setcounter{equation}{0}
    \renewcommand{\theequation}{B.\arabic{equation}}
	\section{Lower bound of the positional precision of astrometric observations}\label{2app}
Diffraction of light and photon statistics limit the positional precision of astrometric observations, and the positional precision for a diffraction-limited image is \citep{lindegren1978photoelectric, van2013astrometry, lindegren2013high}:
          \begin{equation}
             \label{app1}
             \sigma_{lb}= \frac{1}{\pi} \frac{\lambda}{D} \frac{1}{SNR}\ rad \,,
          \end{equation}
          where $\lambda$ is the wavelength of the observed photon; $D$ is the aperture of the telescope; $SNR$ is the signal-to-noise ratio (SNR) in the image, which is given by \citep{howell2006handbook}:
          \begin{equation}
            \label{appp2}
            SNR = \frac{S_{star}}{\sqrt{S_{star}+n_{pix}\times (S_{sky} + S_{dark} + \sigma_{readout}^{2} + (G \times \sigma_f)^2)}}\,,
         \end{equation}
          where $S_{star}$ is the total number of electrons collected from the target star; $n_{pix}$ is the number of pixels under consideration for the SNR calculation; $S_{sky}$ is the total number of electrons per pixel from the background or sky; $S_{dark}$ is the total number of dark current electrons per pixel; $\sigma_{readout}^{2}$ is the total number of electrons per pixel resulting from the read noise; $G$ is the gain of the CCD (in electrons/ADU); $\sigma_f$ is the medium error of the digital-to-analog conversion noise (ADU). 
          \par
          Eq.~(\ref{app1}) is a reference quantity that serves as a lower bound on the positional precision and can be computed without specifying the centroiding algorithm \citep{lindegren2013high}. However, in practice, the lower bound of the positional precision is difficult to reach. Firstly, the sampling process of the detector needs to satisfy the sampling theorem, and when the angular resolution of the telescope pixels is larger than $\frac{\lambda}{2D}$, the positional precision will degrade during the sampling process \citep{lindegren2013high}. For CSST, the lower bound of the positional precision after the theoretical degradation is shown in Fig.~\ref{mngcwc}. Secondly, the centroiding algorithm will also affect the positional precision, and from estimation theory, it can be deduced that only a good centroiding algorithm may come close to this reference lower bound \citep{lindegren2013high}.
          
  \end{appendices}

\end{document}